%% file: main.tex
\documentclass[aps,prd,twocolumn,showpacs,amsmath,amssymb]{revtex4-1}
\usepackage[colorlinks=true,urlcolor=blue]{hyperref}
\usepackage{epsfig}
\usepackage{graphicx}
\usepackage{dcolumn}
\usepackage{bm}
\usepackage{overpic}
\usepackage{subfigure}
\usepackage{float}
\usepackage{color}
\usepackage{amsmath}
\usepackage{mathcomp}
\usepackage{units}
\usepackage{overpic}
\usepackage[T1]{fontenc}
\usepackage[utf8]{inputenc}
\usepackage{multirow}
\usepackage{placeins}
\newcommand{\BESIIIorcid}[1]{\href{https://orcid.org/#1}{\hspace*{0.1em}\raisebox{-0.45ex}{\includegraphics[width=1em]{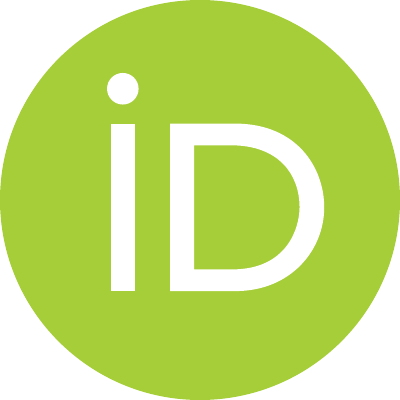}}}} 
\newcommand{\GeV}{\rm GeV}

\newcommand{\gevc}{\,\unit{GeV}/c}
\newcommand{\gevcc}{\,\unit{GeV}/c^2}

\begin{document}
\parskip=5pt plus 1pt minus 1pt
\title{ \boldmath Measurement of the cross sections for $e^+e^- \to p \bar{n} \pi^-$ at center of mass energies from 3.510 to 4.946~$\GeV$}
\author{
  \input{authorlist_2026-04-21}
  }
  \begin{abstract}
    Using a data sample corresponding to an integrated luminosity of 43.8~fb$^{-1}$ 
    collected with the \mbox{BESIII} detector at 44 center-of-mass energy points between 
    3.510 and 4.946~GeV, the Born cross sections for the process 
    $e^+e^- \to p \bar{n} \pi^-$ and its charge-conjugated process are measured.  
    The energy dependence of the dressed cross sections is investigated to search for 
    possible contributions from vector charmonium and charmonium-like states. 
    No evidence for contributions from the $\psi(3770)$, $Y(4230)$, $Y(4360)$, 
    $\psi(4415)$, $Y(4500)$, or $Y(4710)$ resonances is observed. 
    The upper limits at the 90\% confidence level are determined for the products 
    of the electronic partial widths and the branching fractions for the decays of 
    these resonances into the $p \bar{n} \pi^- + \mathrm{c.c.}$ final state.
  \end{abstract}

  \maketitle
  \section{Introduction}\label{sec:introduction}
  Studies of exclusive hadronic processes in $e^+ e^-$ annihilation provide an important 
  testing ground for Quantum Chromodynamics (QCD). Baryonic final states are 
  particularly valuable because they provide access to time-like baryon form factors 
  and possible contributions from exotic hadrons beyond the conventional quark 
  model~\cite{Wang:2025dur}. The process $e^+e^- \to p \bar{n} \pi^-$ is of interest 
  because it involves both baryon-antibaryon and meson production, providing access to 
  the study of intermediate resonances in $e^+e^-$ annihilation. 

  In recent years, numerous charmonium-like hadrons, regarded as candidates for exotic 
  states, have been discovered in $e^+e^-$ annihilation at center-of-mass (c.m.) 
  energies ($\sqrt{s}$) between 3.7 and 
  4.8~$\GeV$~\cite{BES:2001ckj, BESIII:2014rja, BESIII:2023cmv}. While these states 
  have been studied primarily in mesonic channels, their couplings to baryonic final 
  states remain largely unexplored.

  Previous \mbox{BESIII} measurements of the Born cross sections for 
  $e^+e^- \to p \bar{p} \pi^0$ are consistent with perturbative QCD expectations 
  and show no significant resonant 
  structures~\cite{BESIII:2014lpf,BESIII:2017qwj,Wang:2022fwa}. However, theoretical 
  models predict that the virtual photon in $e^+ e^-$ annihilation couples directly 
  to charged particles, making $\pi^\pm$ production more favorable, whereas $\pi^0$ 
  production proceeds only via suppressed neutral mixing~\cite{Gakh:2022fad}. 
  This motivates further exploration of the $p n \pi$ channels. 
  Due to the expected larger production rates, these channels serve as excellent 
  complementary modes to measure the energy-dependent cross sections and search 
  for potential vector exotic states.

  In this paper, the Born cross sections for the process $e^+e^- \to p \bar{n} \pi^- $ 
  are measured at $\sqrt{s}$ between 3.510 and 4.946~$\GeV$. The energy dependence of 
  the dressed cross sections is investigated to search for possible contributions 
  from vector resonances. Charge-conjugate processes are implied throughout this paper 
  unless otherwise specified.

  \section{DETECTOR AND data SAMPLES}\label{sec:BESIII}
  The \mbox{BESIII} detector~\cite{BESIII:2009fln} records symmetric $e^+e^-$ collisions 
  provided by the BEPCII storage ring~\cite{Yu:2016cof} at $\sqrt{s}$ ranging from 
  1.84 to 4.95~$\GeV$, with a peak luminosity of 
  $1.1 \times 10^{33}\;\text{cm}^{-2}\text{s}^{-1}$ achieved at 
  $\sqrt{s} = 3.773\;\GeV$. \mbox{BESIII} has collected large data samples in this 
  energy region~\cite{BESIII:2020nme}. The cylindrical core of the \mbox{BESIII} 
  detector covers 93\% of the full solid angle and consists of a helium-based 
  multilayer drift chamber~(MDC), a time-of-flight system~(TOF), and a CsI(Tl) 
  electromagnetic calorimeter~(EMC), all enclosed in a superconducting solenoidal 
  magnet providing a 1.0~T magnetic field. The solenoid is supported by an octagonal 
  flux-return yoke with resistive plate counter muon identification modules interleaved 
  with steel. The charged-particle momentum resolution at $1~{\rm GeV}/c$ is $0.5\%$. 
  The EMC measures photon energies with a resolution of $2.5\%$ ($5\%$) at $1$~GeV in 
  the barrel (end cap) region. The time resolution in the plastic scintillator TOF 
  barrel region is 68~ps, while that in the end cap region is 110~ps. A muon chamber 
  system (MUC), consisting of $1000~\text{m}^2$ of resistive plate chambers (RPCs), 
  is arranged in nine layers in the barrel and eight layers in the end caps, and is 
  incorporated into the return yoke of the superconducting magnet. The end cap TOF 
  system was upgraded in 2015 using multigap resistive plate chamber technology, 
  providing a time resolution of 60~ps, which benefits 82.4\% of the data used in 
  this analysis~\cite{Li:2017jpg}.

  Monte Carlo (MC) simulated samples produced with a {\sc geant4}-based software 
  package~\cite{GEANT4:2002zbu}, which includes the geometric description of 
  the \mbox{BESIII} detector and the detector response, are used to determine detection 
  efficiencies and to estimate backgrounds. The simulation models the beam energy spread
  and initial state radiation (ISR) in the $e^+e^-$ annihilations using the {\sc kkmc} 
  generator~\cite{Jadach:2000ir}.  

  The inclusive MC sample includes the production of $D\bar{D}$ 
  pairs (including quantum coherence for the neutral $D$ channels), the non-$D\bar{D}$ 
  decays of the $\psi(3770)$, the ISR production of the $J/\psi$ and $\psi(3686)$ states, 
  and the continuum processes incorporated in {\sc kkmc}. In addition, 
  Bhabha scattering~\cite{CarloniCalame:2000pz,CarloniCalame:2003yt}, 
  $e^+e^- \to \mu^+ \mu^-$~\cite{CarloniCalame:2000pz,CarloniCalame:2003yt}, 
  and $e^+e^- \to \tau^+ \tau^-$~\cite{Jadach:2000ir} events are also simulated to 
  account for QED backgrounds. All particle decays, except for continuum processes 
  generated with {\sc kkmc}, are modelled using {\sc evtgen}~\cite{Lange:2001uf}, 
  with branching fractions taken from the 
  Particle Data Group (PDG)~\cite{ParticleDataGroup:2024cfk} when available, or 
  otherwise estimated using {\sc lundcharm}~\cite{Chen:2000tv}. Final state 
  radiation~(FSR) from charged final state particles is incorporated using 
  the {\sc photos} package~\cite{Barberio:1990ms}. This analysis uses 43.8~fb$^{-1}$ 
  of data collected at 44 different energy points from 3.510 to 
  4.946~GeV~\cite{Ablikim:2013ntc, BESIII:2015qfd, BESIII:2024lbn, BESIII:2022dxl, BESIII:2022ulv}. 
  The integrated luminosity at each energy point exceeds 100~pb$^{-1}$.

  \section{EVENT SELECTION AND BACKGROUND STUDY}\label{sec:selection}

  The signal process $e^+e^- \to p \bar{n} \pi^-$ contains a proton, an anti-neutron, and a negatively charged pion in the final state, while its charge-conjugated process contains an anti-proton, a neutron, and a positively charged pion. Candidate events are required to contain exactly two charged tracks with zero net charge. Charged tracks detected in the MDC are required to be within a polar angle ($\theta$) range of $|\cos\theta|<0.93$, where $\theta$ is defined with respect to the $z$-axis, the symmetry axis of the MDC. For each track, the distance $|V_{z}|$ of closest approach along the $z$-axis to the interaction point must be less than 10\,cm, and the distance $|V_{xy}|$ in the transverse plane must be less than 1\,cm.

  In order to identify charged particles, a likelihood-based particle 
  identification (PID) method is used, which combines the energy loss (d$E$/d$x$) 
  measurement in the MDC, the time-of-flight from the TOF, and the electromagnetic 
  shower information from the EMC. Pion candidates are required to have a pion 
  likelihood, calculated using the MDC and TOF information, larger than those for the 
  proton and kaon hypotheses and greater than 0.001. For proton candidates with momentum 
  below 0.8~$\gevc$, the likelihood is calculated using information from the MDC, TOF, 
  and EMC, while only the TOF and EMC information are used for higher-momentum 
  candidates. Proton candidates are required to have a proton likelihood larger 
  than those for the pion, kaon, and electron hypotheses and greater than 0.001.

  To suppress potential backgrounds from Bhabha scattering, the $E/p$ values for proton 
  and $\pi^{\pm}$ candidates are required to be less than 0.8, where $E/p$ is defined 
  as the ratio of the energy deposited in the EMC to the momentum measured in the MDC 
  for charged particles. In the process $e^+e^- \to p \bar{n} \pi^-$, the distribution 
  of the reconstructed vertex position in the transverse plane shows a distinct 
  enhancement at a radial distance of approximately 3~cm from the IP, which is 
  attributed to beam-related interactions with the beam pipe. To suppress this 
  background, events with $R_{xy} >2$~cm are rejected, where 
  $R_{xy} = \sqrt{R_x^2 + R_y^2}$ denotes the radial distance of the reconstructed 
  vertex from the IP in the transverse plane. To suppress the background from cosmic 
  rays, channel-dependent requirements are applied. In the $e^+e^- \to p \bar{n} \pi^-$ 
  channel, the $R_{xy}$ requirement used to suppress beam-related background also 
  removes a significant fraction of the cosmic rays. The remaining cosmic-ray 
  contamination is further suppressed by requiring proton candidates to have a 
  penetration depth in the MUC of less than 40~cm. In contrast, for the 
  $e^+e^- \to \bar{p} n \pi^+$ channel, the angle between the anti-proton and the 
  pion in the $e^+ e^-$ c.m. frame is required to be less than 3.05~rad.

  To reconstruct the four-momentum of the unmeasured neutron (anti-neutron), a 
  one-constraint (1C) kinematic fit is performed by enforcing energy-momentum 
  conservation, where the mass of the missing particle is constrained to the nominal 
  neutron mass. To identify signal candidates, the recoil mass against the selected 
  $p\pi$ system is defined as
  \begin{equation}
    RM_{p\pi} = \sqrt{(\sqrt{s} - E_{p\pi})^2 - \left| \vec{p}_{p\pi} \right|^{2}},
  \end{equation}
  where $E_{p\pi}$ and $\vec{p}_{p\pi}$ are the energy and momentum of the selected candidates in the $e^+ e^-$ c.m. frame.

\begin{figure*}[htbp]
    \centering
    \begin{overpic}[width=0.48\linewidth]{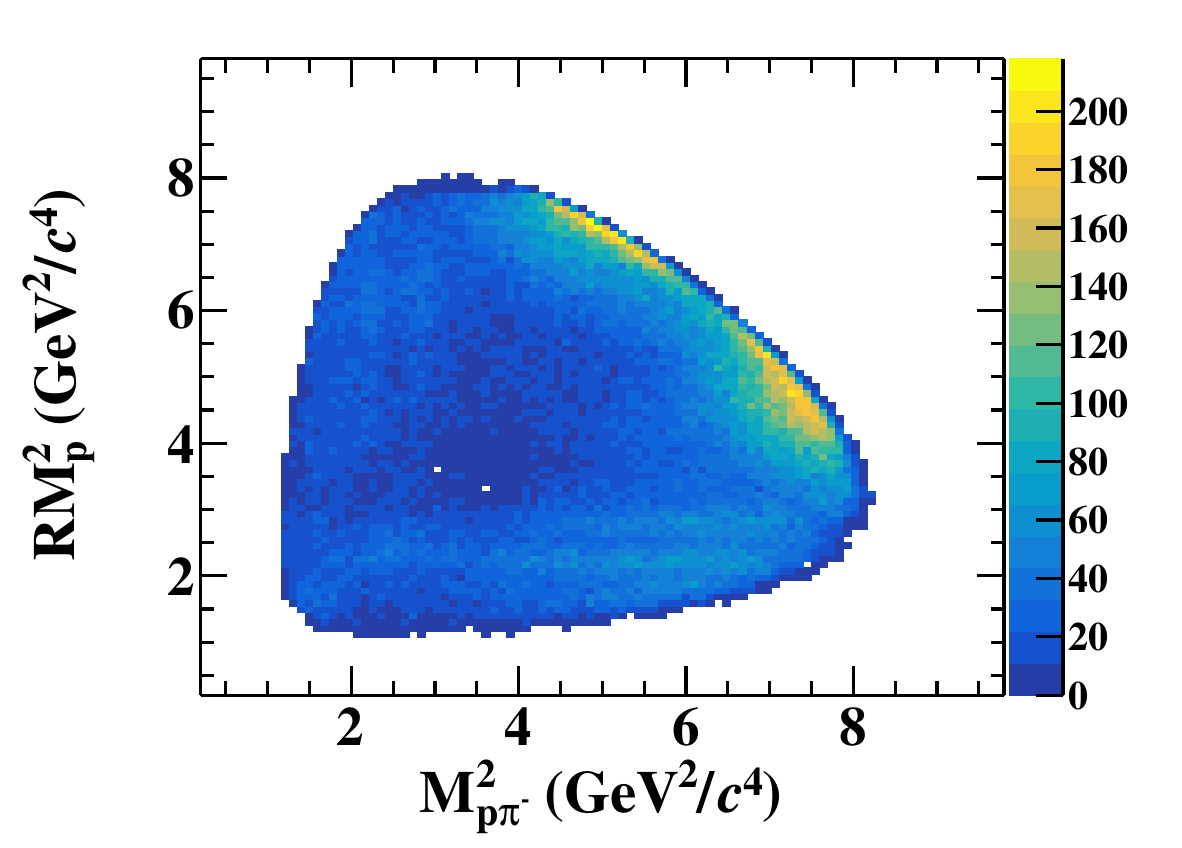}
      \put(58,59){\fontsize{8pt}{8pt}\selectfont\bfseries (a)}
    \end{overpic}
    \begin{overpic}[width=0.48\linewidth]{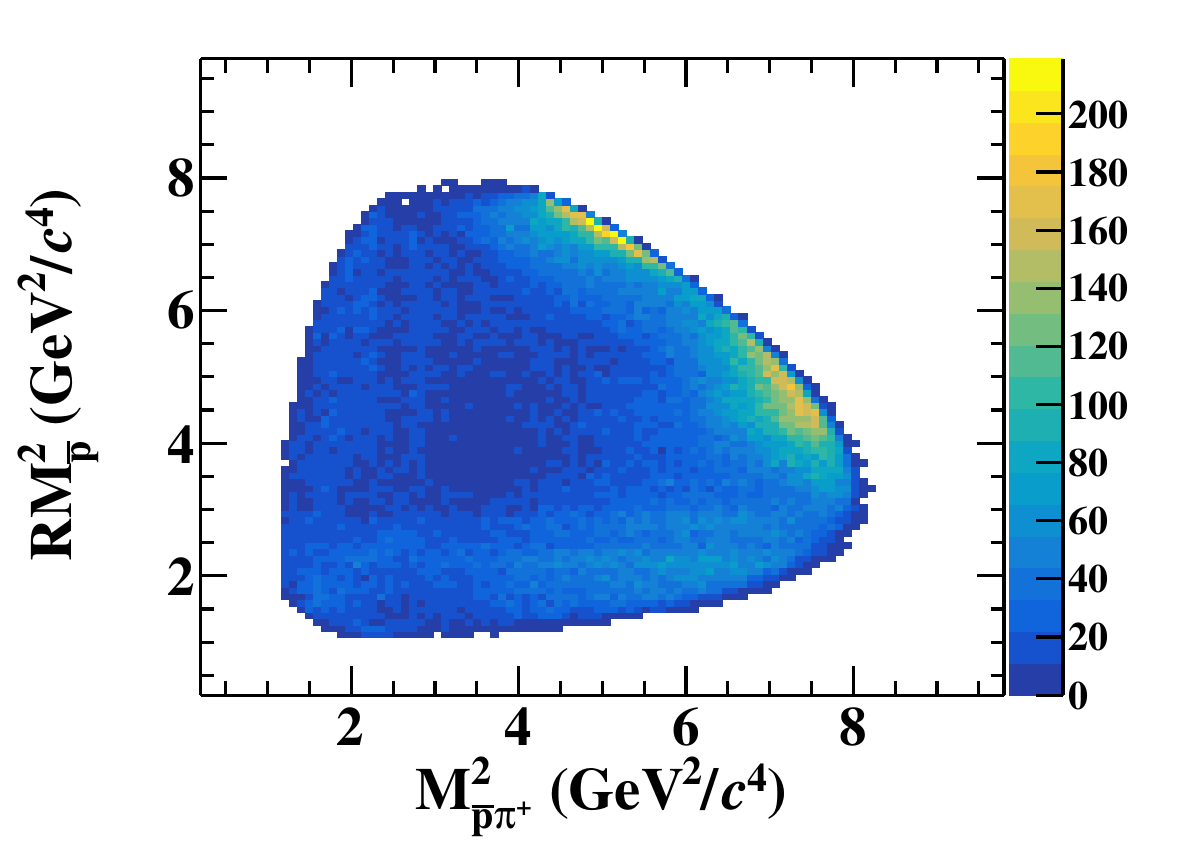}
      \put(58,59){\fontsize{8pt}{8pt}\selectfont\bfseries (b)}
    \end{overpic}
    \begin{overpic}[width=0.48\linewidth]{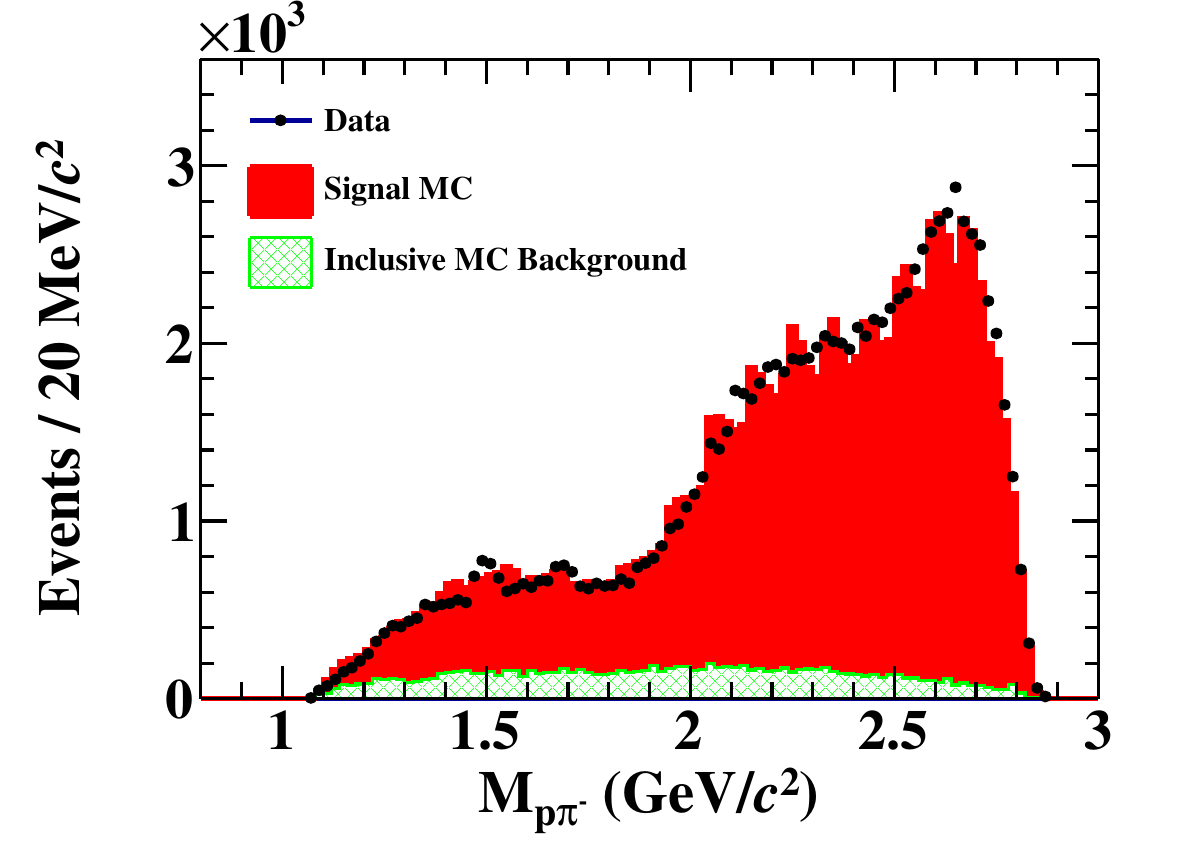}
      \put(58,59){\fontsize{8pt}{8pt}\selectfont\bfseries (c)}
    \end{overpic}
    \begin{overpic}[width=0.48\linewidth]{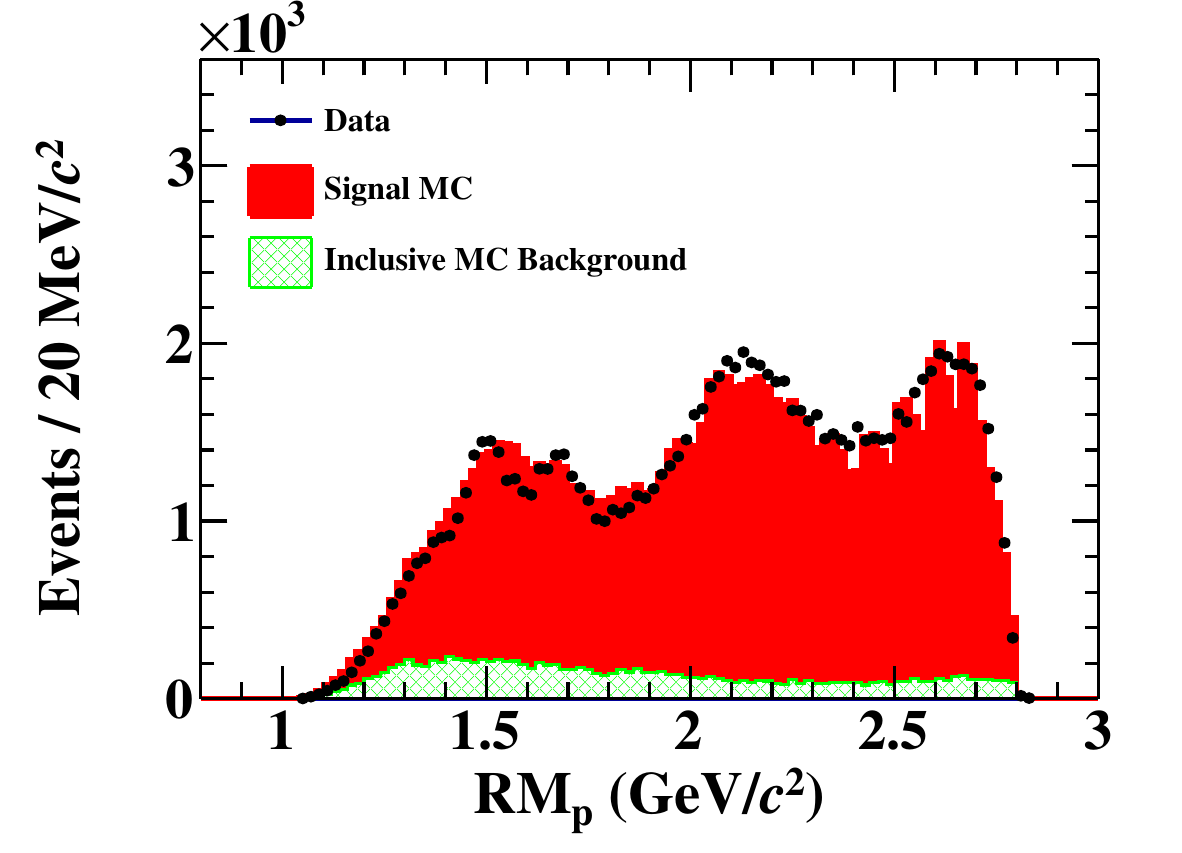}
      \put(58,59){\fontsize{8pt}{8pt}\selectfont\bfseries (d)}
    \end{overpic}
    \begin{overpic}[width=0.48\linewidth]{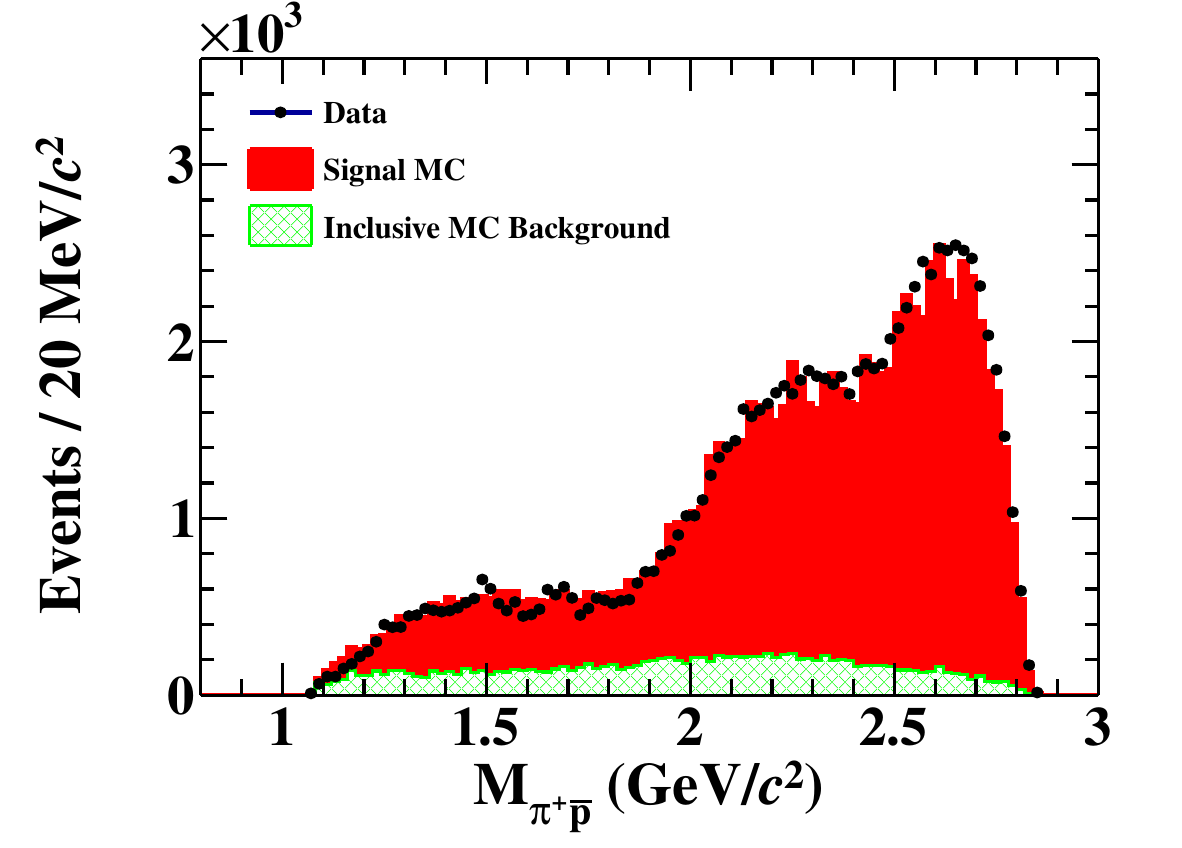}
      \put(58,59){\fontsize{8pt}{8pt}\selectfont\bfseries (e)}
    \end{overpic}
    \begin{overpic}[width=0.48\linewidth]{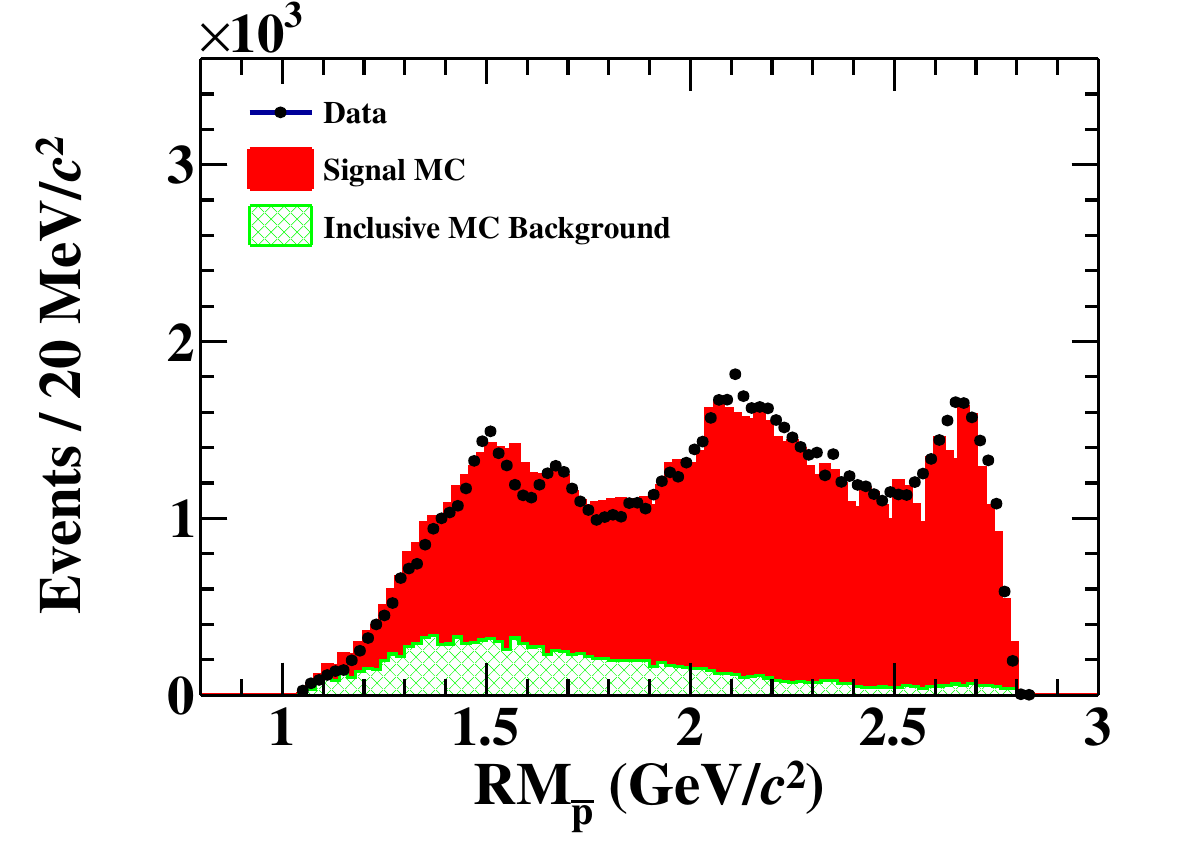}
      \put(58,59){\fontsize{8pt}{8pt}\selectfont\bfseries (f)}
    \end{overpic}
    \caption{(a) Dalitz plot of $M^2_{p \pi^-}$ versus $RM^2_p$; (b) Dalitz plot of 
    $M^2_{\bar{p} \pi^+}$ versus $RM^2_{\bar{p}}$; (c) distribution of $M_{p \pi^-}$; 
    (d) distribution of $RM_p$; (e) distribution of $M_{\bar{p} \pi^+}$; and (f) 
    distribution of $RM_{\bar{p}}$, for all candidate events passing the event 
    selection. For panels (c)--(f), black dots with error bars represent data, 
    red histograms correspond to the signal MC, and the green hatched area with 
    a line contour shows the background estimated from the inclusive MC.}
    \label{fig:phsp}
  \end{figure*}

  The Dalitz plots of the selected events at $\sqrt{s}$ = 3.773 $\GeV$ are shown in the 
  top row of Fig.~\ref{fig:phsp}. The signal MC samples are generated according to the 
  distributions observed in these Dalitz plots. The invariant mass ($M$) distributions 
  of the combined proton-pion ($p\pi$) system and the recoil 
  mass ($RM$) of $p$ or $\bar{p}$ are shown in the bottom two rows of Fig.~\ref{fig:phsp}.

  The backgrounds for $e^+e^- \to p \bar{n} \pi^-$ are studied using the inclusive MC 
  sample at $\sqrt{s}$ = 3.773 $\GeV$. After applying all event selection criteria, 
  the inclusive MC study, complemented by a topology analysis of the backgrounds using 
  the TopoAna package~\cite{Zhou:2020ksj}, indicates that the remaining background 
  primarily originates from the continuum process 
  $e^+ e^- \to \pi^+ \pi^-(k\gamma^{\mathrm{I}})$. Here, $k\gamma^{\mathrm{I}}$ 
  represents one or more photons radiated from the initial state (ISR), 
  where $k \geq 1$. Background contributions from 
  $D\bar{D}$, $\gamma^{\mathrm{I}} J/\psi$, $\gamma^{\mathrm{I}} \psi(2S)$, and QED 
  processes have also been studied. The $p \pi$ recoil mass spectrum shows no evidence 
  of peaking structures in the signal region, and the background distribution can 
  be well described by a Chebyshev polynomial.

  \section{Born cross-section measurement}
  \label{subsec:result}
  \subsection{\label{sec:level2}Extraction of signal yields}
\begin{figure*}[htbp]
    \centering
    \includegraphics[width=0.5\linewidth]{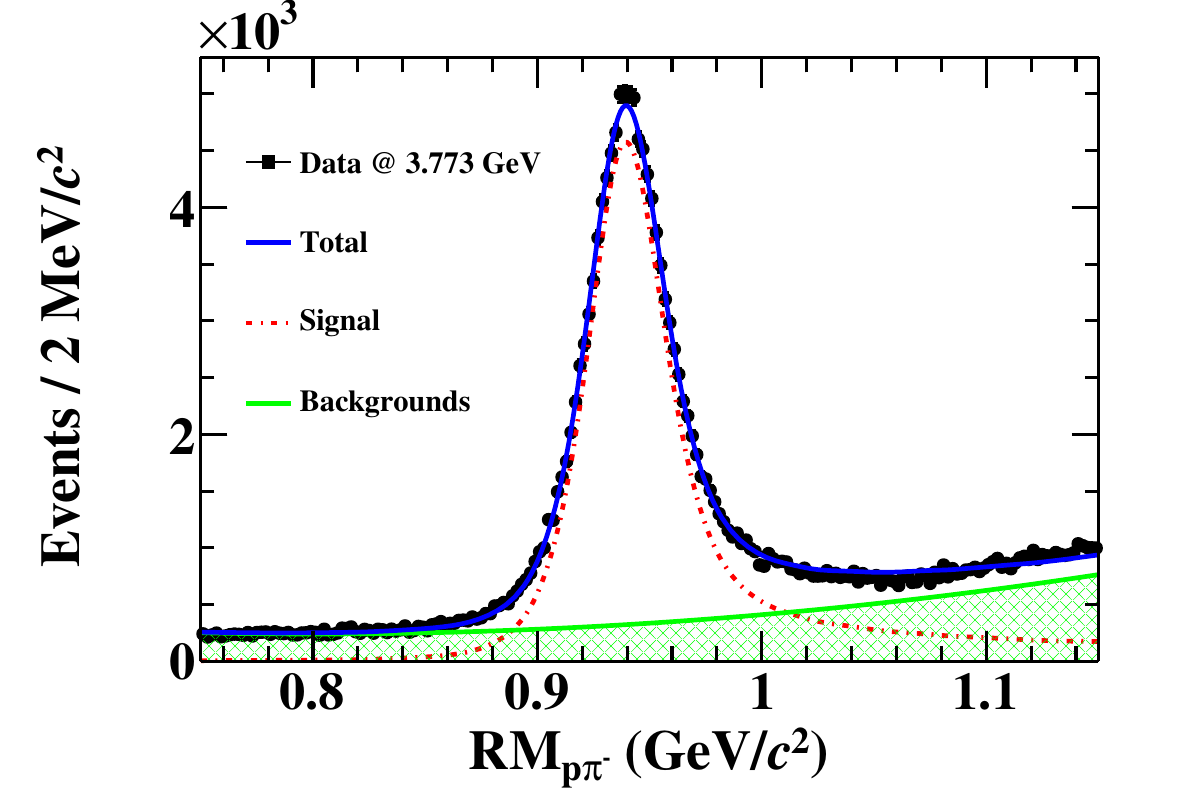}%
    \includegraphics[width=0.5\linewidth]{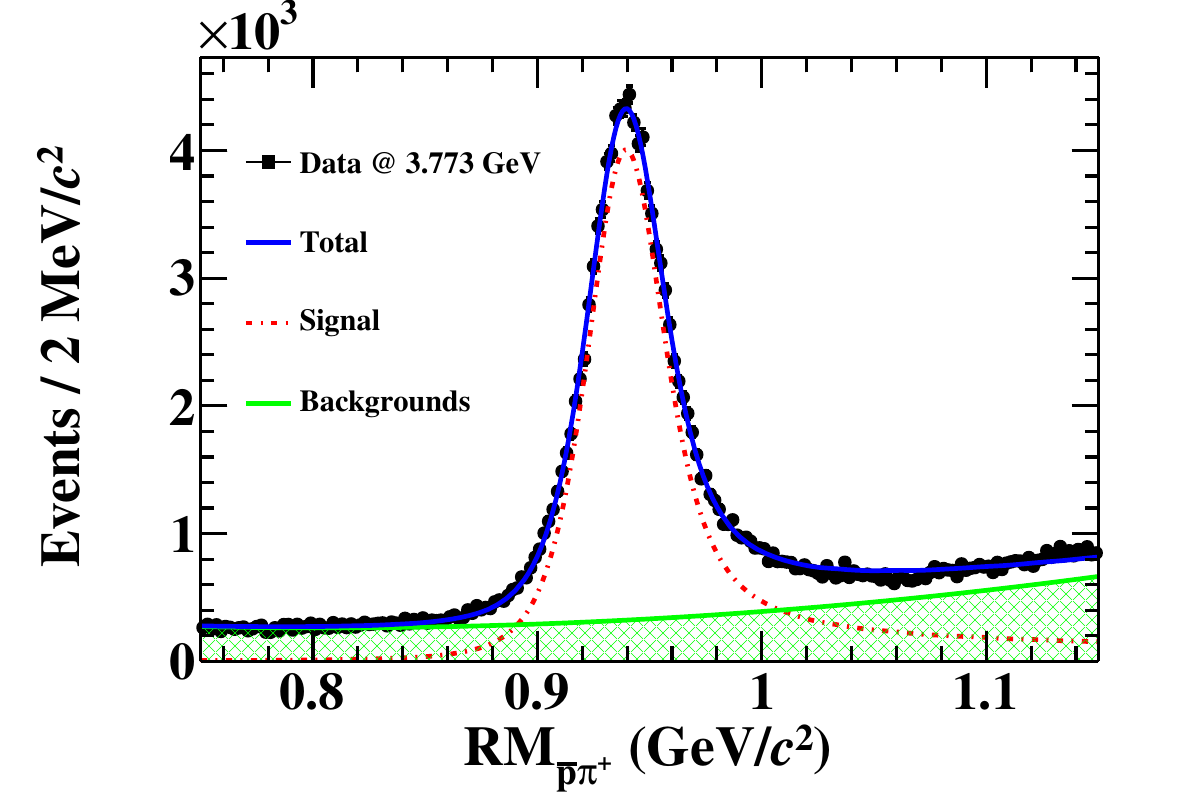}\\
    \caption{Results of the fits to $RM_{p\pi^-}$ and $RM_{\bar{p}\pi^+}$ at 
    $\sqrt{s}$ = 3.773~$\GeV$. The dots with error bars represent the data. The solid 
    blue line shows the result of the fit. The dashed red line indicates the fitted 
    signal, while the green hatched area with a line contour corresponds to the fitted 
    background.}
    \label{fig_fit}
  \end{figure*}

  The signal yields at each energy point are determined using an unbinned 
  maximum-likelihood fit to the $p\pi$ recoil mass spectrum, which is calculated 
  using the measured pre-fit momenta. The signal is described by the MC-simulated 
  distribution convolved with a Gaussian function to account for the resolution 
  difference between data and MC. The background is described by a second-order 
  Chebyshev polynomial. The fit results for the $e^+e^- \to p \bar{n} \pi^-$ and 
  $e^+e^- \to \bar{p} n \pi^+$ channels at $\sqrt{s} = 3.773$~\GeV\ are shown in 
  Fig.~\ref{fig_fit}. The signal yields for all analyzed energy points between 
  $\sqrt{s} = 3.510$ and $4.946$~$\GeV$ are listed in Table~\ref{result_data_pnbarpi} 
  and Table~\ref{result_data_pbarnpi} for the $p \bar{n} \pi^-$ and $\bar{p} n \pi^+$ 
  channels, respectively.

  \subsection{\label{sec:level2}Determination of Born cross sections}

  Taking the $e^+e^- \to p \bar{n} \pi^-$ channel as an example, the Born cross 
  sections are calculated using
  \begin{equation}\label{eq_Born}
    \sigma^{B}(e^+e^- \to p \bar{n} \pi^-) = \frac{N^{\mathrm{obs}}}{\epsilon\cdot\mathcal{L}\cdot(1+\delta)\cdot\frac{1}{|1-\Pi|^2}},
  \end{equation}
  where $N^{\mathrm{obs}}$ is the signal yield, $\mathcal{L}$ is the integrated 
  luminosity, and $\epsilon$ is the selection efficiency. The factor $(1+\delta)$ is 
  the radiative correction factor, which is taken from a QED calculation using the 
  measured cross-section line shape as input in an iterative 
  procedure~\cite{Sun:2020ehv}, while $\frac{1}{|1-\Pi|^2}$ is the vacuum 
  polarization (VP) factor including both leptonic and hadronic 
  contributions~\cite{Jegerlehner:2011ti}. In this analysis, a single set of signal MC 
  samples including ISR effects is generated. An iterative MC-weighting method is then 
  applied to correct the detection efficiencies while simultaneously obtaining the ISR 
  correction factors. The iteration continues until convergence is achieved, which is 
  defined by requiring the relative difference in $(1+\delta) \cdot \epsilon$ between 
  two successive iterations to be less than 0.1\%. The final values of the detection 
  efficiency, ISR correction factor, and Born cross sections are obtained through this 
  procedure and are listed in Tables~\ref{result_data_pnbarpi} and \ref{result_data_pbarnpi}.

  \begin{table*}[htbp]    
    \centering    
    \setlength{\tabcolsep}{12pt} 
    \caption{Summary of $e^+e^-\to p\bar{n}\pi^-$ at each energy point. The table lists 
    the center-of-mass energy $\sqrt{s}$, integrated luminosity $\mathcal{L}$, observed 
    signal yields $N^{\mathrm{obs}}$, detection efficiency $\epsilon$, radiative 
    correction factor $1+\delta$, vacuum polarization factor $\frac{1}{|1-\Pi|^2}$, 
    and the measured Born cross section $\sigma^{B}$. For $\sigma^{B}$, the first 
    uncertainties are statistical and the second are systematic. A detailed breakdown 
    of the systematic uncertainties is summarized in Table~\ref{sytem_err_pnbarpi}.}    
    \begin{tabular}{c|c|c|c|c|c|c}    
      \hline\hline    
      $\sqrt{s}$~($\GeV$) & $\mathcal{L}$~(pb$^{-1}$) & $N^{\mathrm{obs}}$ & $\epsilon~$($\%$) & 1 + $\delta$ & $\frac{1}{|1-\Pi|^2}$ & $\sigma^{B}$~(pb)\\    
      \hline
      3.510 & 405.70 & 4794.2 $\pm$ 88.5 & 43.02 $\pm$ 0.22 & 1.39 & 1.04 & 18.99 $\pm$ 0.35 $\pm$ 0.87\\    
      3.554 & 129.40 & 1317.8 $\pm$ 46.5 & 43.19 $\pm$ 0.33 & 1.41 & 1.04 & 16.10 $\pm$ 0.57 $\pm$ 0.89\\    
      3.650 & 410.00 & 3390.8 $\pm$ 75.3 & 40.16 $\pm$ 0.22 & 1.46 & 1.02 & 13.86 $\pm$ 0.31 $\pm$ 0.56\\    
      3.768 & 415.80 & 2905.9 $\pm$ 73.8 & 38.30 $\pm$ 0.23 & 1.53 & 1.05 & 11.32 $\pm$ 0.29 $\pm$ 0.43\\    
      3.773 & $20.27 \times 10^3$ & 135721.7 $\pm$ 533.4 & 37.81 $\pm$ 0.04 & 1.52 & 1.06 & 11.03 $\pm$ 0.04 $\pm$ 0.41\\    
      3.780 & 410.00 & 2951.8 $\pm$ 73.8 & 38.30$ \pm$ 0.22 & 1.54 & 1.06 & 11.56 $\pm$ 0.29 $\pm$ 0.46\\    
      3.872 & 219.20 & 1237.7 $\pm$ 48.0 & 36.19 $\pm$ 0.28 & 1.58 & 1.05 & 9.39 $\pm$ 0.36 $\pm$ 0.41\\    
      4.009 & 482.00 & 2153.3 $\pm$ 70.4 & 33.68 $\pm$ 0.21 & 1.86 & 1.04 & 6.85 $\pm$ 0.22 $\pm$ 0.27\\    
      4.130 & 401.50 & 1462.3 $\pm$ 58.1 & 32.07 $\pm$ 0.24 & 1.75 & 1.05 & 6.15 $\pm$ 0.24 $\pm$ 0.23\\    
      4.160 & 408.70 & 1351.1 $\pm$ 58.2 & 31.40 $\pm$ 0.24 & 1.78 & 1.05 & 5.62 $\pm$ 0.24 $\pm$ 0.26\\    
      4.180 & 3189.00 & 10700.2 $\pm$ 149.1 & 31.69 $\pm$ 0.08 & 1.79 & 1.05 & 5.60 $\pm$ 0.08 $\pm$ 0.28\\    
      4.190 & 526.70 & 1718.9 $\pm$ 59.6 & 31.76 $\pm$ 0.19 & 1.81 & 1.06 & 5.37 $\pm$ 0.19 $\pm$ 0.35\\    
      4.200 & 526.00 & 1644.9 $\pm$ 59.4 & 31.39 $\pm$ 0.19 & 1.81 & 1.06 & 5.21 $\pm$ 0.19 $\pm$ 0.25\\    
      4.210 & 517.10 & 1576.3 $\pm$ 59.1 & 30.96 $\pm$ 0.19 & 1.82 & 1.06 & 5.11 $\pm$ 0.19 $\pm$ 0.26\\    
      4.220 & 514.60 & 1623.0 $\pm$ 60.1 & 31.10 $\pm$ 0.19 & 1.83 & 1.06 & 5.26 $\pm$ 0.19 $\pm$ 0.28\\    
      4.230 & 1056.40 & 3467.1 $\pm$ 85.3 & 32.56 $\pm$ 0.14 & 1.83 & 1.06 & 5.22 $\pm$ 0.13 $\pm$ 0.22\\    
      4.237 & 530.30 & 1601.8 $\pm$ 58.6 & 31.87 $\pm$ 0.19 & 1.84 & 1.06 & 4.89 $\pm$ 0.18 $\pm$ 0.21\\    
      4.246 & 538.10 & 1597.8 $\pm$ 57.8 & 31.64 $\pm$ 0.19 & 1.84 & 1.06 & 4.82 $\pm$ 0.17 $\pm$ 0.18\\    
      4.260 & 828.40 & 2409.1 $\pm$ 74.1 & 31.78 $\pm$ 0.17 & 1.86 & 1.05 & 4.66 $\pm$ 0.14 $\pm$ 0.20\\    
      4.270 & 531.10 & 1437.1 $\pm$ 55.6 & 31.16 $\pm$ 0.19 & 1.87 & 1.05 & 4.42 $\pm$ 0.17 $\pm$ 0.22\\    
      4.280 & 175.70 & 480.7 $\pm$ 32.2 & 30.63 $\pm$ 0.34 & 1.87 & 1.05 & 4.53 $\pm$ 0.30 $\pm$ 0.24\\    
      4.290 & 502.40 & 1431.5 $\pm$ 54.9 & 31.63 $\pm$ 0.20 & 1.89 & 1.05 & 4.53 $\pm$ 0.17 $\pm$ 0.19\\    
      4.315 & 501.20 & 1385.0 $\pm$ 55.0 & 31.10 $\pm$ 0.20 & 1.91 & 1.05 & 4.42 $\pm$ 0.18 $\pm$ 0.18\\    
      4.340 & 505.00 & 1278.4 $\pm$ 54.1 & 31.02 $\pm$ 0.20 & 1.93 & 1.05 & 4.03 $\pm$ 0.17 $\pm$ 0.16\\    
      4.360 & 543.90 & 1454.4 $\pm$ 58.3 & 30.05 $\pm$ 0.20 & 1.94 & 1.05 & 4.36 $\pm$ 0.17 $\pm$ 0.20\\    
      4.380 & 522.70 & 1208.9 $\pm$ 53.4 & 30.18 $\pm$ 0.19 & 1.97 & 1.05 & 3.70 $\pm$ 0.16 $\pm$ 0.18\\    
      4.400 & 507.80 & 1204.4 $\pm$ 54.1 & 29.89 $\pm$ 0.20 & 1.98 & 1.05 & 3.81 $\pm$ 0.17 $\pm$ 0.17\\
      4.420 & 1043.90 & 2362.4 $\pm$ 77.1 & 29.38 $\pm$ 0.12 & 2.00 & 1.05 & 3.66 $\pm$ 0.12 $\pm$ 0.16\\    
      4.440 & 569.90 & 1322.7 $\pm$ 59.0 & 29.41 $\pm$ 0.20 & 2.02 & 1.05 & 3.71 $\pm$ 0.17 $\pm$ 0.16\\    
      4.470 & 111.09 & 207.8 $\pm$ 24.6 & 28.81 $\pm$ 0.34 & 2.05 & 1.05 & 3.01 $\pm$ 0.36 $\pm$ 0.20\\    
      4.530 & 112.12 & 266.7 $\pm$ 26.7 & 27.77 $\pm$ 0.34 & 2.09 & 1.05 & 3.88 $\pm$ 0.39 $\pm$ 0.28\\    
      4.600 & 586.90 & 1068.9 $\pm$ 55.4 & 26.57 $\pm$ 0.19 & 2.16 & 1.05 & 3.00 $\pm$ 0.16 $\pm$ 0.18\\    
      4.610 & 103.83 & 186.3 $\pm$ 23.2 & 26.15 $\pm$ 0.33 & 2.18 & 1.05 & 2.98 $\pm$ 0.37 $\pm$ 0.22\\    
      4.620 & 521.52 & 834.7 $\pm$ 50.1 & 25.70 $\pm$ 0.19 & 2.20 & 1.05 & 2.68 $\pm$ 0.16 $\pm$ 0.18\\    
      4.640 & 522.41 & 935.7 $\pm$ 51.1 & 25.69 $\pm$ 0.19 & 2.21 & 1.05 & 2.99 $\pm$ 0.16 $\pm$ 0.22\\    
      4.660 & 529.63 & 796.8 $\pm$ 49.3 & 25.00 $\pm$ 0.19 & 2.23 & 1.05 & 2.56 $\pm$ 0.16 $\pm$ 0.18\\    
      4.680 & 1669.31 & 2517.6 $\pm$ 88.0 & 24.86 $\pm$ 0.11 & 2.26 & 1.05 & 2.54 $\pm$ 0.09 $\pm$ 0.18\\    
      4.700 & 536.45 & 802.8 $\pm$ 49.5 & 24.72 $\pm$ 0.19 & 2.29 & 1.05 & 2.51 $\pm$ 0.15 $\pm$ 0.19\\    
      4.740 & 164.27 & 210.9 $\pm$ 29.2 & 23.89 $\pm$ 0.33 & 2.33 & 1.05 & 2.19 $\pm$ 0.30 $\pm$ 0.19\\    
      4.750 & 367.21 & 436.0 $\pm$ 41.0 & 23.63 $\pm$ 0.27 & 2.35 & 1.05 & 2.03 $\pm$ 0.19 $\pm$ 0.15\\    
      4.780 & 512.78 & 616.7 $\pm$ 47.0 & 23.67 $\pm$ 0.20 & 2.37 & 1.06 & 2.03 $\pm$ 0.15 $\pm$ 0.16\\    
      4.840 & 527.29 & 515.8 $\pm$ 58.3 & 22.54 $\pm$ 0.18 & 2.46 & 1.06 & 1.67 $\pm$ 0.19 $\pm$ 0.13\\    
      4.914 & 208.11 & 182.0 $\pm$ 27.9 & 21.79 $\pm$ 0.33 & 2.53 & 1.06 & 1.50 $\pm$ 0.23 $\pm$ 0.17\\    
      4.946 & 160.37 & 149.7 $\pm$ 26.5 & 21.43 $\pm$ 0.33 & 2.58 & 1.06 & 1.60 $\pm$ 0.28 $\pm$ 0.15\\
      \hline\hline
    \end{tabular}       
    \label{result_data_pnbarpi}
  \end{table*}

  \begin{table*}   
    \centering   
    \setlength{\tabcolsep}{12pt} 
    \caption{Summary of $e^+e^-\to \bar{p} n \pi^+$ at each energy point. The table 
    lists the center-of-mass energy $\sqrt{s}$, integrated luminosity $\mathcal{L}$, 
    observed signal yields $N^{\mathrm{obs}}$, detection efficiency $\epsilon$, 
    radiative correction factor $1+\delta$, vacuum polarization factor 
    $\frac{1}{|1-\Pi|^2}$, and the measured Born cross section $\sigma^{B}$. 
    For $\sigma^{B}$, the first uncertainties are statistical and the second ones 
    are systematic. A detailed breakdown of the systematic uncertainties is summarized 
    in Table~\ref{sytem_err_pbarnpi}.}.
    \begin{tabular}{c|c|c|c|c|c|c}    
      \hline\hline    
      $\sqrt{s}$~($\GeV$)& $\mathcal{L}$~(pb$^{-1}$) & $N^{\mathrm{obs}}$ & $\epsilon$~($\%$) & 1 + $\delta$ & $\frac{1}{|1-\Pi|^2}$ & $\sigma^{B}$~(pb)\\    
      \hline
      3.510 & 405.70 & 4273.3 $\pm$ 81.5 & 38.49 $\pm$ 0.22 & 1.40 & 1.04 & 18.76 $\pm$ 0.36 $\pm$ 1.05\\    
      3.554 & 129.40 & 1278.8 $\pm$ 45.2 & 38.16 $\pm$ 0.33 & 1.42 & 1.04 & 17.54 $\pm$ 0.62 $\pm$ 0.91\\    
      3.650 & 410.00 & 2991.1 $\pm$ 72.5 & 36.25 $\pm$ 0.22 & 1.47 & 1.02 & 13.42 $\pm$ 0.33 $\pm$ 0.73\\    
      3.768 & 415.80 & 2430.6 $\pm$ 70.9 & 34.30 $\pm$ 0.22 & 1.54 & 1.05 & 10.46 $\pm$ 0.31 $\pm$ 0.56\\    
      3.773 & $20.27 \times 10^3$ & 119310.1 $\pm$ 512.8 & 33.90 $\pm$ 0.04 & 1.54 & 1.06 & 10.70 $\pm$ 0.05 $\pm$ 0.52\\    
      3.780 & 410.00 & 2427.4 $\pm$ 68.9 & 34.28 $\pm$ 0.22 & 1.55 & 1.06 & 10.50 $\pm$ 0.30 $\pm$ 0.55\\    
      3.872 & 219.20 & 1078.3 $\pm$ 46.9 & 32.88 $\pm$ 0.28 & 1.60 & 1.05 & 8.90 $\pm$ 0.39 $\pm$ 0.44\\    
      4.009 & 482.00 & 2037.4 $\pm$ 76.8 & 31.23 $\pm$ 0.20 & 1.88 & 1.04 & 6.89 $\pm$ 0.26 $\pm$ 0.37\\    
      4.130 & 401.50 & 1326.3 $\pm$ 59.6 & 29.35 $\pm$ 0.23 & 1.78 & 1.05 & 6.00 $\pm$ 0.27 $\pm$ 0.29\\    
      4.160 & 408.70 & 1230.7 $\pm$ 59.7 & 28.89 $\pm$ 0.23 & 1.81 & 1.05 & 5.47 $\pm$ 0.27 $\pm$ 0.27\\    
      4.180 & 3189.00 & 9733.5 $\pm$ 146.1 & 29.25 $\pm$ 0.07 & 1.82 & 1.05 & 5.43 $\pm$ 0.08 $\pm$ 0.31\\    
      4.190 & 526.70 & 1553.1 $\pm$ 59.7 & 29.02 $\pm$ 0.19 & 1.84 & 1.06 & 5.23 $\pm$ 0.20 $\pm$ 0.30\\    
      4.200 & 526.00 & 1548.9 $\pm$ 60.0 & 28.99 $\pm$ 0.19 & 1.84 & 1.06 & 5.22 $\pm$ 0.20 $\pm$ 0.28\\    
      4.210 & 517.10 & 1480.2 $\pm$ 59.3 & 28.48 $\pm$ 0.19 & 1.85 & 1.06 & 5.13 $\pm$ 0.21 $\pm$ 0.27\\    
      4.220 & 514.60 & 1280.2 $\pm$ 56.1 & 28.52 $\pm$ 0.19 & 1.86 & 1.06 & 4.45 $\pm$ 0.20 $\pm$ 0.27\\    
      4.230 & 1056.40 & 3478.9 $\pm$ 95.5 & 30.36 $\pm$ 0.16 & 1.86 & 1.06 & 5.53 $\pm$ 0.15 $\pm$ 0.30\\    
      4.237 & 530.30 & 1387.8 $\pm$ 56.2 & 29.47 $\pm$ 0.19 & 1.87 & 1.06 & 4.50 $\pm$ 0.18 $\pm$ 0.29\\    
      4.246 & 538.10 & 1411.2 $\pm$ 56.9 & 29.06 $\pm$ 0.19 & 1.88 & 1.06 & 4.55 $\pm$ 0.18 $\pm$ 0.24\\    
      4.260 & 828.40 & 2371.2 $\pm$ 76.7 & 29.69 $\pm$ 0.16 & 1.90 & 1.05 & 4.83 $\pm$ 0.16 $\pm$ 0.25\\    
      4.270 & 531.10 & 1389.7 $\pm$ 57.8 & 28.88$\pm$ 0.19 & 1.90 & 1.05 & 4.53 $\pm$ 0.19 $\pm$ 0.24\\    
      4.280 & 175.70 & 439.4 $\pm$ 32.1 & 28.35 $\pm$ 0.33 & 1.91 & 1.05 & 4.39 $\pm$ 0.32 $\pm$ 0.24\\    
      4.290 & 502.40 & 1256.5 $\pm$ 57.5 & 29.60 $\pm$ 0.20 & 1.92 & 1.05 & 4.17 $\pm$ 0.19 $\pm$ 0.22\\    
      4.315 & 501.20 & 1221.9 $\pm$ 54.9 & 29.10 $\pm$ 0.20 & 1.95 & 1.05 & 4.09 $\pm$ 0.18 $\pm$ 0.26\\    
      4.340 & 505.00 & 1145.6 $\pm$ 55.0 & 28.78 $\pm$ 0.20 & 1.96 & 1.05 & 3.82 $\pm$ 0.18 $\pm$ 0.19\\    
      4.360 & 543.90 & 1261.9 $\pm$ 59.9 & 28.26 $\pm$ 0.19 & 1.98 & 1.05 & 3.94 $\pm$ 0.19 $\pm$ 0.20\\    
      4.380 & 522.70 & 1091.0 $\pm$ 54.4 & 28.07 $\pm$ 0.19 & 2.01 & 1.05 & 3.52 $\pm$ 0.18 $\pm$ 0.19\\    
      4.400 & 507.80 & 1063.9 $\pm$ 53.4 & 27.82 $\pm$ 0.20 & 2.02 & 1.05 & 3.55 $\pm$ 0.18 $\pm$ 0.21\\    
      4.420 & 1043.90 & 2106.7 $\pm$ 79.5 & 27.60 $\pm$ 0.12 & 2.04 & 1.05 & 3.40 $\pm$ 0.13 $\pm$ 0.19\\    
      4.440 & 569.90 & 1221.2 $\pm$ 58.6 & 27.33 $\pm$ 0.19 & 2.06 & 1.05 & 3.61 $\pm$ 0.17 $\pm$ 0.19\\    
      4.470 & 111.09 & 224.0 $\pm$ 25.3 & 27.32 $\pm$ 0.33 & 2.09 & 1.05 & 3.35 $\pm$ 0.38 $\pm$ 0.21\\    
      4.530 & 112.12 & 212.0 $\pm$ 25.9 & 26.41 $\pm$ 0.33 & 2.14 & 1.05 & 3.17 $\pm$ 0.39 $\pm$ 0.20\\    
      4.600 & 586.90 & 981.8 $\pm$ 59.9 & 25.23 $\pm$ 0.19 & 2.21 & 1.05 & 2.84 $\pm$ 0.17 $\pm$ 0.19\\    
      4.610 & 103.83 & 146.0 $\pm$ 23.0 & 24.71 $\pm$ 0.33 & 2.23 & 1.05 & 2.42 $\pm$ 0.38 $\pm$ 0.17\\    
      4.620 & 521.52 & 786.2 $\pm$ 53.0 & 24.17 $\pm$ 0.18 & 2.25 & 1.05 & 2.63 $\pm$ 0.18 $\pm$ 0.24\\    
      4.640 & 522.41 & 797.3 $\pm$ 54.7 & 24.15 $\pm$ 0.19 & 2.26 & 1.05 & 2.65 $\pm$ 0.18 $\pm$ 0.19\\    
      4.660 & 529.63 & 647.3 $\pm$ 51.7 & 23.73 $\pm$ 0.18 & 2.29 & 1.05 & 2.14 $\pm$ 0.17 $\pm$ 0.17\\    
      4.680 & 1669.31 & 2446.6 $\pm$ 94.1 & 23.52 $\pm$ 0.11 & 2.32 & 1.05 & 2.55 $\pm$ 0.10 $\pm$ 0.20\\    
      4.700 & 536.45 & 651.4 $\pm$ 51.5 & 23.36 $\pm$ 0.18 & 2.34 & 1.05 & 2.10 $\pm$ 0.17 $\pm$ 0.19\\    
      4.740 & 164.27 & 231.9 $\pm$ 29.3 & 22.60 $\pm$ 0.33 & 2.39 & 1.05 & 2.48 $\pm$ 0.31 $\pm$ 0.19\\    
      4.750 & 367.21 & 469.5 $\pm$ 41.8 & 22.48 $\pm$ 0.27 & 2.41 & 1.05 & 2.24 $\pm$ 0.20 $\pm$ 0.17\\    
      4.780 & 512.78 & 512.9 $\pm$ 48.8 & 22.26 $\pm$ 0.20 & 2.44 & 1.06 & 1.75 $\pm$ 0.17 $\pm$ 0.14\\    
      4.840 & 527.29 & 472.1 $\pm$ 50.7 & 21.37 $\pm$ 0.18 & 2.53 & 1.06 & 1.57 $\pm$ 0.17 $\pm$ 0.19\\    
      4.914 & 208.11 & 193.9 $\pm$ 32.1 & 20.60 $\pm$ 0.32 & 2.61 & 1.06 & 1.64 $\pm$ 0.27 $\pm$ 0.15\\    
      4.946 & 160.37 & 86.8 $\pm$ 26.9 & 20.30 $\pm$ 0.32 & 2.66 & 1.06 & 0.95 $\pm$ 0.29 $\pm$ 0.09\\
      \hline\hline
    \end{tabular}     
    \label{result_data_pbarnpi}
  \end{table*}

  \begin{table*}[htbp]    
    \centering    
    \caption{Systematic uncertainties (in~\%) and their sources for each energy point 
    on the Born cross sections measurement for $e^+e^- \to p \bar{n} \pi^-$. 
    Sources of systematic uncertainty that are found to be negligible are not listed 
    in this table.} 
    \begin{tabular}{c|c|c|c|c|c|c|c|c|c|c}    
      \hline\hline    
      $\sqrt{s}$ (GeV) & $\mathcal{L}$ & Tracking &  PID ($p$) & PID ($\pi^{-}$) & 
      Signal shape & \begin{tabular}{@{}c@{}}Background \\ shape\end{tabular} & 
        Fit range & MC model & ISR correction & Total\\
        \hline
        3.510 & 1.0 & 2.0 & 1.2 & 1.2 & 0.1 & 1.9 & 0.9 & 1.8 & 2.3 & 4.6 \\
        3.554 & 1.0 & 2.0 & 1.2 & 1.2 & 0.1 & 3.7 & 1.3 & 1.8 & 2.1 & 5.5 \\
        3.650 & 1.0 & 2.0 & 1.3 & 1.2 & 0.2 & 1.0 & 0.8 & 1.8 & 1.8 & 4.0 \\
        3.768 & 1.0 & 2.0 & 1.3 & 1.2 & 0.7 & 0.6 & 0.4 & 1.8 & 1.5 & 3.8 \\
        3.773 & 1.0 & 2.0 & 1.3 & 1.2 & 0.8 & 1.2 & 0.3 & 1.0 & 1.5 & 3.7 \\
        3.780 & 1.0 & 2.0 & 1.3 & 1.2 & 0.0 & 0.5 & 1.3 & 1.8 & 1.5 & 3.9 \\
        3.872 & 1.0 & 2.0 & 1.4 & 1.2 & 0.0 & 2.0 & 1.3 & 1.8 & 1.2 & 4.3 \\
        4.009 & 1.0 & 2.0 & 1.5 & 1.2 & 0.0 & 1.5 & 0.4 & 1.8 & 0.9 & 3.9 \\
        4.130 & 1.0 & 2.0 & 1.6 & 1.2 & 0.3 & 0.6 & 0.8 & 1.8 & 0.6 & 3.7 \\
        4.160 & 1.0 & 2.0 & 1.7 & 1.2 & 0.8 & 2.8 & 0.3 & 1.8 & 0.5 & 4.6 \\
        4.180 & 1.0 & 2.0 & 1.7 & 1.2 & 0.0 & 3.1 & 1.6 & 1.8 & 0.6 & 5.0 \\
        4.190 & 1.0 & 2.0 & 1.7 & 1.2 & 0.0 & 4.9 & 2.3 & 1.8 & 0.6 & 6.5 \\
        4.200 & 1.0 & 2.0 & 1.7 & 1.2 & 0.5 & 1.9 & 2.4 & 1.8 & 0.6 & 4.7 \\
        4.210 & 1.0 & 2.0 & 1.7 & 1.2 & 0.0 & 3.4 & 1.3 & 1.8 & 0.6 & 5.1 \\
        4.220 & 1.0 & 2.0 & 1.8 & 1.2 & 0.1 & 3.0 & 2.7 & 1.8 & 0.6 & 5.4 \\
        4.230 & 1.0 & 2.0 & 1.8 & 1.2 & 0.2 & 2.0 & 1.0 & 1.8 & 0.6 & 4.2 \\
        4.237 & 1.0 & 2.0 & 1.8 & 1.2 & 0.0 & 1.6 & 1.8 & 1.8 & 0.6 & 4.4 \\
        4.246 & 1.0 & 2.0 & 1.8 & 1.2 & 0.0 & 0.2 & 0.6 & 1.8 & 0.6 & 3.7 \\
        4.260 & 1.0 & 2.0 & 1.8 & 1.2 & 0.0 & 1.9 & 0.8 & 1.8 & 0.6 & 4.2 \\
        4.270 & 1.0 & 2.0 & 1.8 & 1.2 & 0.0 & 3.0 & 1.8 & 1.8 & 0.6 & 5.0 \\
        4.280 & 1.0 & 2.0 & 1.8 & 1.2 & 0.1 & 1.6 & 3.5 & 1.8 & 0.6 & 5.3 \\
        4.290 & 1.0 & 2.0 & 1.8 & 1.3 & 0.1 & 0.9 & 1.9 & 1.8 & 0.6 & 4.2 \\
        4.315 & 1.0 & 2.0 & 1.8 & 1.3 & 0.1 & 0.3 & 1.9 & 1.8 & 0.6 & 4.1 \\
        4.340 & 1.0 & 2.0 & 1.8 & 1.3 & 0.0 & 1.0 & 0.9 & 1.8 & 0.6 & 3.9 \\
        4.360 & 1.0 & 2.0 & 1.8 & 1.3 & 0.1 & 1.9 & 2.0 & 1.8 & 0.6 & 4.6 \\
        4.380 & 1.0 & 2.0 & 1.9 & 1.3 & 0.0 & 3.0 & 1.2 & 1.8 & 0.6 & 4.9 \\
        4.400 & 1.0 & 2.0 & 1.9 & 1.3 & 0.0 & 2.3 & 0.8 & 1.8 & 0.6 & 4.4 \\
        4.420 & 1.0 & 2.0 & 1.9 & 1.4 & 0.1 & 1.9 & 1.0 & 1.8 & 0.6 & 4.4 \\
        4.440 & 1.0 & 2.0 & 1.9 & 1.7 & 0.1 & 0.5 & 1.8 & 1.8 & 0.6 & 4.3 \\
        4.470 & 1.0 & 2.0 & 2.0 & 1.9 & 0.1 & 5.2 & 1.3 & 1.8 & 0.6 & 6.7 \\
        4.530 & 1.0 & 2.0 & 2.0 & 3.5 & 0.5 & 3.5 & 3.7 & 1.8 & 0.6 & 7.1 \\
        4.600 & 1.0 & 2.0 & 2.0 & 4.0 & 0.0 & 2.3 & 2.0 & 1.8 & 0.6 & 6.2 \\
        4.610 & 1.0 & 2.0 & 2.0 & 4.0 & 3.3 & 2.9 & 2.3 & 1.8 & 0.6 & 7.4 \\
        4.620 & 1.0 & 2.0 & 2.0 & 4.8 & 0.0 & 1.5 & 2.3 & 1.8 & 0.6 & 6.5 \\
        4.640 & 1.0 & 2.0 & 2.0 & 4.9 & 0.1 & 1.2 & 3.8 & 1.8 & 0.6 & 7.2 \\
        4.660 & 1.0 & 2.0 & 2.0 & 5.7 & 0.8 & 0.9 & 1.2 & 1.8 & 0.6 & 6.9 \\
        4.680 & 1.0 & 2.0 & 2.0 & 5.7 & 0.5 & 0.9 & 1.4 & 1.8 & 0.6 & 6.9 \\
        4.700 & 1.0 & 2.0 & 2.0 & 5.8 & 0.5 & 1.7 & 2.2 & 1.8 & 0.6 & 7.4 \\
        4.740 & 1.0 & 2.0 & 2.0 & 5.9 & 5.1 & 0.5 & 2.2 & 1.8 & 0.6 & 8.8 \\
        4.750 & 1.0 & 2.0 & 2.0 & 5.9 & 0.2 & 1.0 & 2.0 & 1.8 & 0.6 & 7.2 \\
        4.780 & 1.0 & 2.0 & 2.0 & 6.0 & 1.6 & 0.9 & 2.7 & 1.8 & 0.6 & 7.6 \\
        4.840 & 1.0 & 2.0 & 2.0 & 6.0 & 0.1 & 1.7 & 3.4 & 1.8 & 0.6 & 7.9 \\
        4.914 & 1.0 & 2.0 & 2.0 & 7.1 & 2.4 & 6.2 & 4.6 & 1.8 & 0.6 & 11.3 \\
        4.946 & 1.0 & 2.0 & 2.1 & 7.3 & 1.3 & 0.5 & 4.5 & 1.8 & 0.6 & 9.4 \\
        \hline\hline    
    \end{tabular}      
    \label{sytem_err_pnbarpi}
  \end{table*}

  \begin{table*}[htbp]    
    \centering    
    \caption{Systematic uncertainties (in~\%) and their sources for each energy point 
    on the Born cross sections measurement for $e^+e^- \to \bar{p} n \pi^+$. 
    Sources of systematic uncertainty that are found to be negligible are not listed 
    in this table.}   
    \begin{tabular}{c|c|c|c|c|c|c|c|c|c|c}    
      \hline\hline    
      $\sqrt{s}$ (GeV) & $\mathcal{L}$ & Tracking &  PID ($\bar{p}$) & PID ($\pi^{+}$) 
      & Signal shape & \begin{tabular}{@{}c@{}}Background \\ shape\end{tabular} & 
        Fit range & MC model & ISR correction & Total\\     
        \hline
        3.510 & 1.0 & 2.0 & 2.8 & 1.2 & 0.1 & 2.9 & 1.4 & 1.8 & 1.7 & 5.6 \\
        3.554 & 1.0 & 2.0 & 2.9 & 1.2 & 0.0 & 1.9 & 1.7 & 1.8 & 1.5 & 5.2 \\
        3.650 & 1.0 & 2.0 & 2.9 & 1.2 & 0.4 & 2.2 & 2.1 & 1.8 & 1.2 & 5.4 \\
        3.768 & 1.0 & 2.0 & 3.0 & 1.2 & 2.4 & 0.3 & 1.7 & 1.8 & 0.8 & 5.3 \\
        3.773 & 1.0 & 2.0 & 3.0 & 1.2 & 1.9 & 1.7 & 0.5 & 1.0 & 0.8 & 4.9 \\
        3.780 & 1.0 & 2.0 & 3.0 & 1.2 & 1.9 & 2.0 & 0.8 & 1.8 & 0.7 & 5.3 \\
        3.872 & 1.0 & 2.0 & 3.1 & 1.2 & 1.0 & 1.0 & 1.7 & 1.8 & 0.6 & 4.9 \\
        4.009 & 1.0 & 2.0 & 3.2 & 1.2 & 0.1 & 2.2 & 1.9 & 1.8 & 0.6 & 5.4 \\
        4.130 & 1.0 & 2.0 & 3.4 & 1.2 & 0.1 & 0.0 & 1.3 & 1.8 & 0.6 & 4.8 \\
        4.160 & 1.0 & 2.0 & 3.4 & 1.2 & 0.5 & 0.5 & 1.6 & 1.8 & 0.6 & 5.0 \\
        4.180 & 1.0 & 2.0 & 3.4 & 1.2 & 0.1 & 2.8 & 1.6 & 1.8 & 0.6 & 5.7 \\
        4.190 & 1.0 & 2.0 & 3.4 & 1.2 & 2.2 & 2.4 & 0.2 & 1.8 & 0.6 & 5.7 \\
        4.200 & 1.0 & 2.0 & 3.4 & 1.2 & 1.8 & 1.9 & 1.1 & 1.8 & 0.6 & 5.4 \\
        4.210 & 1.0 & 2.0 & 3.5 & 1.2 & 1.4 & 1.8 & 0.2 & 1.8 & 0.6 & 5.2 \\
        4.220 & 1.0 & 2.0 & 3.5 & 1.2 & 2.8 & 2.9 & 0.5 & 1.8 & 0.6 & 6.2 \\
        4.230 & 1.0 & 2.0 & 3.5 & 1.2 & 1.6 & 1.3 & 1.8 & 1.8 & 0.6 & 5.4 \\
        4.237 & 1.0 & 2.0 & 3.5 & 1.2 & 2.9 & 3.1 & 0.4 & 1.8 & 0.6 & 6.4 \\
        4.246 & 1.0 & 2.0 & 3.5 & 1.2 & 1.1 & 1.4 & 1.5 & 1.8 & 0.6 & 5.2 \\
        4.260 & 1.0 & 2.0 & 3.5 & 1.2 & 1.1 & 1.1 & 1.0 & 1.8 & 0.6 & 5.1 \\
        4.270 & 1.0 & 2.0 & 3.5 & 1.2 & 0.1 & 2.0 & 1.1 & 1.8 & 0.6 & 5.3 \\
        4.280 & 1.0 & 2.0 & 3.5 & 1.2 & 1.5 & 1.8 & 1.5 & 1.8 & 0.6 & 5.5 \\
        4.290 & 1.0 & 2.0 & 3.6 & 1.3 & 1.4 & 1.4 & 0.7 & 1.8 & 0.6 & 5.2 \\
        4.315 & 1.0 & 2.0 & 3.6 & 1.3 & 2.4 & 2.4 & 2.3 & 1.8 & 0.6 & 6.3 \\
        4.340 & 1.0 & 2.0 & 3.6 & 1.3 & 0.0 & 0.0 & 1.4 & 1.8 & 0.6 & 5.0 \\
        4.360 & 1.0 & 2.0 & 3.6 & 1.3 & 0.7 & 1.0 & 1.4 & 1.8 & 0.6 & 5.2 \\
        4.380 & 1.0 & 2.0 & 3.6 & 1.3 & 1.2 & 1.3 & 1.1 & 1.8 & 0.6 & 5.3 \\
        4.400 & 1.0 & 2.0 & 3.7 & 1.3 & 2.0 & 2.0 & 1.8 & 1.8 & 0.6 & 5.9 \\
        4.420 & 1.0 & 2.0 & 3.7 & 1.4 & 1.5 & 1.9 & 0.8 & 1.8 & 0.6 & 5.5 \\
        4.440 & 1.0 & 2.0 & 3.7 & 1.7 & 0.3 & 0.3 & 1.4 & 1.8 & 0.6 & 5.2 \\
        4.470 & 1.0 & 2.0 & 3.7 & 2.0 & 2.0 & 2.1 & 1.9 & 1.8 & 0.6 & 6.1 \\
        4.530 & 1.0 & 2.0 & 3.7 & 3.5 & 0.1 & 1.3 & 1.9 & 1.8 & 0.6 & 6.2 \\
        4.600 & 1.0 & 2.0 & 3.7 & 4.0 & 0.5 & 1.4 & 2.1 & 1.8 & 0.6 & 6.7 \\
        4.610 & 1.0 & 2.0 & 3.7 & 4.0 & 1.2 & 1.6 & 3.1 & 1.8 & 0.6 & 7.2 \\
        4.620 & 1.0 & 2.0 & 3.7 & 4.8 & 5.5 & 0.4 & 3.2 & 1.8 & 0.6 & 9.3 \\
        4.640 & 1.0 & 2.0 & 3.7 & 4.9 & 1.0 & 0.2 & 1.6 & 1.8 & 0.6 & 7.0 \\
        4.660 & 1.0 & 2.0 & 3.7 & 5.7 & 2.6 & 0.9 & 1.9 & 1.8 & 0.6 & 8.1 \\
        4.680 & 1.0 & 2.0 & 3.7 & 5.7 & 1.1 & 0.7 & 1.9 & 1.8 & 0.6 & 7.8 \\
        4.700 & 1.0 & 2.0 & 3.7 & 5.8 & 3.5 & 3.4 & 1.8 & 1.8 & 0.6 & 9.1 \\
        4.740 & 1.0 & 2.0 & 3.7 & 5.9 & 0.7 & 0.6 & 0.8 & 1.8 & 0.6 & 7.6 \\
        4.750 & 1.0 & 2.0 & 3.7 & 5.9 & 0.8 & 0.8 & 1.6 & 1.8 & 0.6 & 7.8 \\
        4.780 & 1.0 & 2.0 & 3.7 & 6.0 & 1.2 & 0.9 & 1.5 & 1.8 & 0.6 & 7.9 \\
        4.840 & 1.0 & 2.0 & 3.7 & 6.0 & 5.7 & 5.8 & 4.7 & 1.8 & 0.6 & 12.1 \\
        4.914 & 1.0 & 2.0 & 3.7 & 7.0 & 1.2 & 1.4 & 3.6 & 1.8 & 0.6 & 9.4 \\
        4.946 & 1.0 & 2.0 & 3.8 & 7.3 & 0.0 & 1.8 & 1.4 & 1.8 & 0.6 & 9.0 \\  
        \hline\hline    
    \end{tabular}     
    \label{sytem_err_pbarnpi}
  \end{table*}

  \section{Systematic Uncertainty}
  \label{subsec:BF}
  The following sources of systematic uncertainty are considered in the cross-section measurement:  
  \begin{itemize}
    \item \textbf{Luminosity:}  
      The integrated luminosity at each energy point is measured using Bhabha 
      scattering events~\cite{Ablikim:2013ntc, BESIII:2015qfd, BESIII:2024lbn, BESIII:2022dxl, BESIII:2022ulv}, with a relative uncertainty of 1.0\%.
    \item \textbf{Tracking:}  
      The tracking efficiency is evaluated using the control sample 
      $e^{+} e^{-} \to p\bar{p}\pi^{+}\pi^{-}$~\cite{BESIII:2019hdp}, 
      and an uncertainty of 1.0\% is assigned to each charged track.
    \item \textbf{PID of proton:}  
      \begin{itemize}
        \item For protons, three selection criteria are considered: the PID likelihood, 
          the $E/p$ ratio, and the penetration depth in the MUC. The corresponding 
          systematic uncertainties are evaluated using a high-purity 
          $e^{+} e^{-} \to p\bar{p}\pi^{+}\pi^{-}$ control sample. Candidate events 
          are required to contain four good charged tracks with zero net charge. 
          These tracks are constrained to a common vertex and identified as pions or 
          antiprotons using the PID requirements described above. A four-constraint (4C) 
          kinematic fit is applied to the selected tracks. Events are retained if 
          $\chi^2 < 10$, which efficiently suppresses background contributions. 
          The background contribution is evaluated with inclusive MC samples and is 
          found to be below 1\%, which is neglected. To account for the momentum 
          coverage of the control sample, a fit is performed to evaluate the systematic 
          uncertainty for protons with momenta exceeding 0.8~GeV/$c$, where modified 
          PID criteria are adopted. The total systematic uncertainty for each energy 
          point, $\delta_{\text{total}}$, is determined by weighting the maximum 
          uncertainty in each momentum interval by its corresponding fractional 
          contribution:
          \begin{equation}
            \delta_{\text{total}} = \sum_{i} \omega_{i} \cdot \delta_{i}^{\text{max}},
          \end{equation}
          here $\omega_{i} = N_{i} / N_{\text{total}}$ represents the fraction of signal 
          events in the $i$-th momentum interval, and $\delta_{i}^{\text{max}}$ is the 
          maximum systematic uncertainty assigned to that interval. 
          Using the selected control samples, $\delta_{i}^{\text{max}}$ is determined 
          from the relative difference in PID efficiencies between 
          data ($\epsilon_{\text{data},i}$) and MC simulation ($\epsilon_{\text{MC},i}$) 
          in each momentum bin, 
          formulated as $\delta_{i}^{\text{max}} = | \epsilon_{\text{data},i} / \epsilon_{\text{MC},i} - 1 |$. 
          Here, the PID efficiencies are evaluated by applying the same proton 
          PID selection criteria as those implemented in the nominal event selection 
          in Sec.~\ref{sec:selection}.     
          The resulting PID uncertainties for protons at each energy point are 
          summarized in Table~\ref{sytem_err_pnbarpi}.
        \item For antiprotons, only the PID likelihood requirement is applied. 
          The same control sample $e^{+} e^{-} \to p\bar{p}\pi^{+}\pi^{-}$ is used. 
          The event selection criteria are identical to those used in the previous 
          systematic uncertainty study. To account for the momentum coverage of the 
          control sample, the same momentum-weighting procedure as described above is 
          adopted. The resulting PID uncertainties for antiprotons at each energy point 
          are listed in Table~\ref{sytem_err_pbarnpi}.
      \end{itemize}   
    \item \textbf{PID of pion:}  The systematic uncertainty associated with charged-pion 
      PID is evaluated using a control sample of $D\bar{D}$ hadronic 
      decays~\cite{BESIII:2018apz}. To account for the momentum coverage of the control 
      sample, the momentum-weighting procedure described above is adopted. Furthermore, 
      a systematic uncertainty of 0.7\% is assigned to the $E/p$ requirement, which is 
      determined using a control sample of $e^+e^- \to \pi^+\pi^-\pi^0$ 
      events~\cite{BESIII:2024okl}. The resulting systematic uncertainties for pion 
      PID are listed in Table~\ref{sytem_err_pnbarpi} and Table~\ref{sytem_err_pbarnpi} 
      for the $p\bar{n}\pi^-$ and $\bar{p}n\pi^+$ channels, respectively.
    \item \textbf{Vertex position ($R_{xy}$) selection:}  
      Since the $R_{xy}$ requirement is identical for all energy points, the data 
      collected at $\sqrt{s} = 3.773$~$\GeV$ are used as the control sample. 
      A Barlow test~\cite{Barlow:2002yb} is performed to quantify the significance 
      of the deviation between the nominal result and results obtained by varying 
      the $R_{xy}$ selection criterion. The Barlow test statistic $\zeta$ is defined as  
      \begin{equation}
        \zeta = \frac{|S_{\mathrm{nominal}} - S_{\mathrm{test}}|}{\sqrt{|\delta^2_{S_{\mathrm{nominal}}} - \delta^2_{S_{\mathrm{test}}}|}},
      \end{equation}
      where $S$ denotes the measured cross section and $\delta_S$ is its statistical 
      uncertainty. A $\zeta$ value close to zero indicates good agreement between the 
      nominal result and results obtained by varying the $R_{xy}$ selection, while a 
      large $\zeta$ suggests a possible systematic discrepancy. The systematic 
      uncertainty due to the vertex-position requirement is estimated by varying the 
      $R_{xy}$ cut from 1.7 to 2.3~cm in steps of 0.1~cm. The largest difference 
      relative to the nominal value (2.0~cm) is assigned as the uncertainty. In all 
      cases, the resulting $\zeta$ remains below 2, indicating that the systematic 
      uncertainty associated with the vertex-position selection is negligible.
    \item \textbf{Cosmic ray background suppression requirement:}  
      To evaluate the potential systematic effect of this requirement, the nominal 
      selection of $\theta_{\bar{p}\pi^+}^{\mathrm{CMS}} < 3.05$~rad is replaced by an 
      alternative selection based on the penetration depth of the charged tracks in the 
      MUC. Both criteria suppress cosmic-ray backgrounds, since cosmic-ray tracks 
      typically penetrate deeply into the MUC and exhibit a back-to-back signature. 
      The efficiency for data and MC is calculated as 
      $\epsilon = N_{\mathrm{after}}/N_{\mathrm{before}}$, where $N_{\mathrm{before}}$ 
      and $N_{\mathrm{after}}$ are the event yields before and after applying the 
      requirement, respectively. The difference between the data and MC efficiencies 
      is taken as the systematic uncertainty. The resulting efficiency change is 0.07\%, 
      which is negligible.  
    \item \textbf{Fit method}
      \begin{itemize}
        \item Signal shape: The nominal signal shape is modeled by an MC-simulated 
          distribution convolved with a Gaussian function, and fitted together with a 
          second-order Chebyshev polynomial to describe the background. To evaluate the 
          systematic uncertainty associated with the signal shape, the Gaussian convolution is removed, and the fit is repeated using the MC shape plus the same second-order 
          Chebyshev polynomial for the background. The difference in the extracted 
          number of signal events between the nominal and alternative fits is taken as 
          the systematic uncertainty, and the resulting uncertainties for each energy 
          point are summarized in Table~\ref{sytem_err_pnbarpi} and 
          Table~\ref{sytem_err_pbarnpi}.
        \item Background shape: In the nominal fit, the background is parameterized 
          with a second-order Chebyshev polynomial. To estimate the associated 
          systematic uncertainty, the background parameterization is replaced with a 
          third-order Chebyshev polynomial, while keeping the signal 
          shape (MC convolved with Gaussian) unchanged. The resulting difference in the 
          number of signal events is assigned as the systematic uncertainty, with the 
          results summarized in Table~\ref{sytem_err_pnbarpi} and Table~\ref{sytem_err_pbarnpi}.
        \item Fit range: The systematic uncertainty associated with the mass-window 
          selection is evaluated using a Barlow test. The nominal mass windows are 
          chosen to be $[0.75, 1.15]$~$\gevcc$, $[0.65, 1.25]$~$\gevcc$, and 
          $[0.60, 1.30]$~$\gevcc$ for data samples at $\sqrt{s} \le 4.160$~GeV, 
          $4.180 \le \sqrt{s} \le 4.220$~GeV, and $4.230 \le \sqrt{s} \le 4.946$~GeV, 
          respectively. The mass window is varied from $-60$ to $+60$~MeV/$c^2$ 
          relative to the nominal window in steps of 20~MeV/$c^2$. 
          The corresponding variations in the measured cross sections are evaluated. 
          Among all tested window sizes, the variation with the largest Barlow 
          significance is selected, and the corresponding difference with respect 
          to the nominal result is assigned as the systematic uncertainty.
          The resulting uncertainties for all energy points are summarized in 
          Table~\ref{sytem_err_pnbarpi} and Table~\ref{sytem_err_pbarnpi}.
      \end{itemize}
    \item \textbf{MC model:} The {\sc diy} generator~\cite{Jadach:2000ir} simulates the 
      signal decay according to an input Dalitz plot, which is constructed from 
      background-subtracted data and corrected for the detection efficiency using 
      phase-space MC. At the energy point of 3.773~$\GeV$, where the statistics are 
      highest, the difference in detection efficiencies between the Dalitz-based and 
      phase-space MC samples is found to be 1.0\%, which is taken as the systematic 
      uncertainty at this energy. To further evaluate the possible dependence on 
      $\sqrt{s}$, similar tests are performed at other energy points. The largest 
      observed difference, 1.8\%, is assigned as the systematic uncertainty for the 
      other energy points, considering their relatively limited statistics.
    \item \textbf{ISR correction:}  
      The uncertainty from the ISR correction consists of two components. First, it is 
      evaluated by sampling the cross sections within their uncertainties. The resulting 
      parameters are used to recalculate the ISR correction factor $f_{\mathrm{ISR}}$ 
      and efficiency $\epsilon$. This procedure is repeated 1000 times, yielding 1000 
      determinations of $f_{\mathrm{ISR}}\cdot\epsilon$. A Gaussian function is fitted 
      to these distributions to extract the mean ($\mu$) and standard 
      deviation ($\sigma$), and the ratio $\sigma/\mu$ is taken as one source of the 
      systematic uncertainty. Second, an additional uncertainty of about 0.5\% is 
      assigned due to the radiative function, as reported in a previous 
      study~\cite{Ping:2013jka}. The resulting uncertainties for each energy point are 
      summarized in Table~\ref{sytem_err_pnbarpi} and Table~\ref{sytem_err_pbarnpi}.
    \item \textbf{VP correction:}
      The VP factor is calculated with an uncertainty of less than 
      0.1\%~\cite{WorkingGrouponRadiativeCorrections:2010bjp}, which is negligible 
      compared with other sources of uncertainty.
  \end{itemize}

  All the systematic uncertainties are summarized in Table~\ref{sytem_err_pnbarpi} and 
  Table~\ref{sytem_err_pbarnpi}, where the total uncertainties are obtained by adding 
  all contributions in quadrature, assuming statistical independence of all sources.
  \begin{figure}  
    \centering  
    \includegraphics[width=1.0\linewidth]{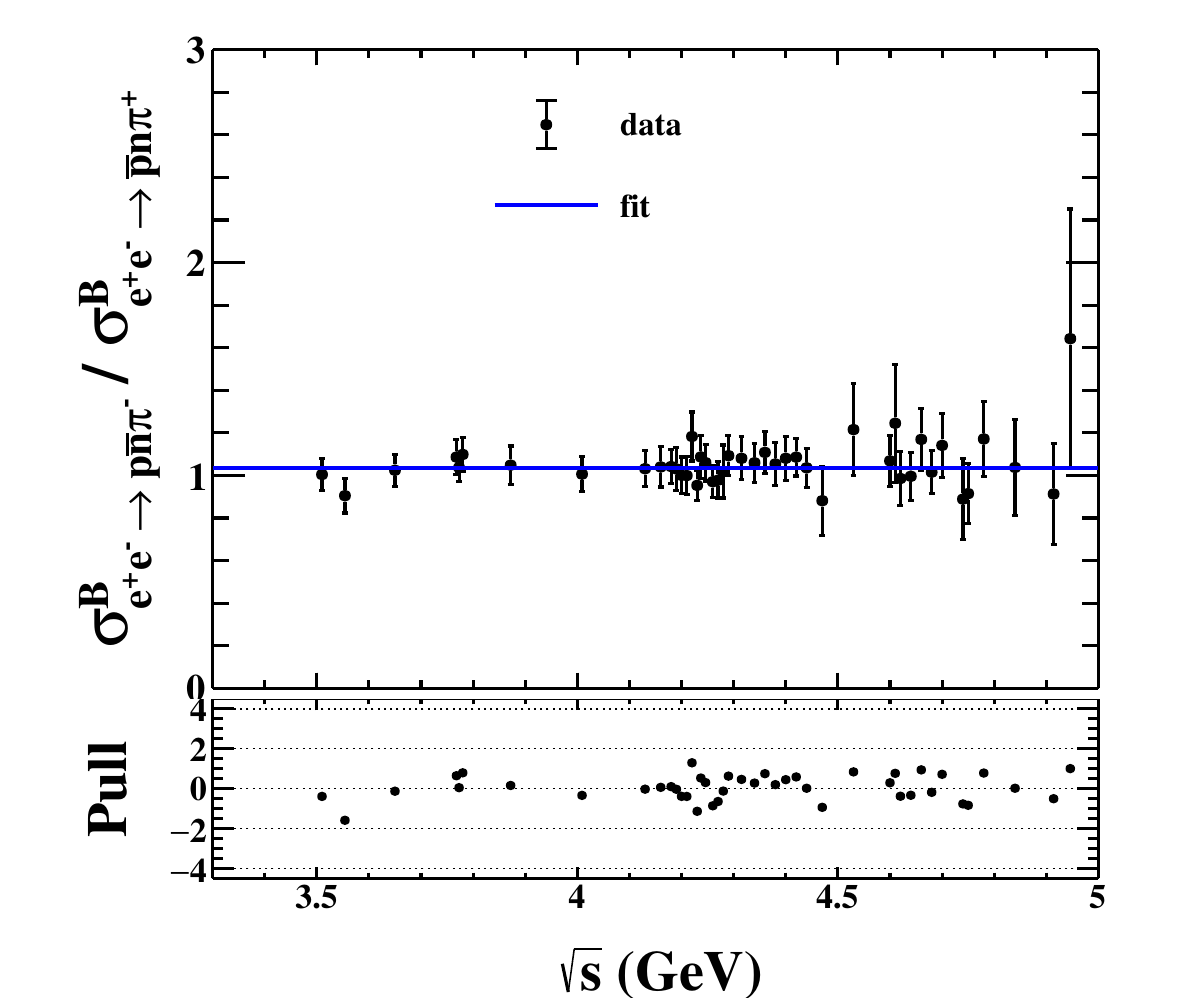} 
    \caption{The ratio of the measured cross sections for 
    $e^{+}e^{-}\to p\bar{n}\pi^{-}$ to $e^{+}e^{-}\to \bar{p}n\pi^{+}$ at different 
    energy points. The black dots with error bars represent the data ratios, and the 
    blue solid line represents the fit result. The bottom panel shows the corresponding 
    pull distribution of the fit.}  
    \label{fig:com_cs_pnpi}
  \end{figure}

  \begin{table*}[htbp]    
    \centering    
    \caption{The fitted resonance parameters for the decay of 
    $e^{+}e^{-}\to p \bar{n} \pi^{-}$ with two solutions (I, II). 
    The reported uncertainties include both statistical and systematic errors.}    
    \resizebox{\textwidth}{!}{    
    \begin{tabular}{c|c|c|c|c|c|c|c}    
      \hline\hline 
      Resonance  & $C$ $(10^5$ $\text{GeV}^{2\lambda}\text{pb})$ & \( \lambda \) & $\phi$ (rad) & $\Gamma_{e^+ e^-} \times \mathcal{B}$ (eV) & Significance ($\sigma$) & $\chi^2$/NDF & $\Gamma_{e^+ e^-} \times \mathcal{B}$ (eV) at the 90\% C.L. \\    
      \hline
      $\psi$(3770)$^I$ & 2.83 $\pm$ 0.46 & 3.55 $\pm$ 0.06 & 6.26 $\pm$ 3.89 & (2.10 $\pm$ 5.14) $\times$ 10$^{-4}$ & \multirow{2}{*}{0.4} & \multirow{2}{*}{30.6/($44-4$)} & 8.69 $\times$ 10$^{-4}$ \\    
      $\psi$(3770)$^{II}$ & 2.84 $\pm$ 0.46 & 3.55 $\pm$ 0.06 & 4.71 $\pm$ 0.02 & (9.47 $\pm$ 0.13) $\times$ 10$^{-1}$ & & & $\ldots$ \\  
      \hline    
      $Y(4230)$$^I$ & 2.70 $\pm$ 0.40 & 3.53 $\pm$ 0.05 & 2.63 $\pm$ 0.91 & (6.16 $\pm$ 11.90) $\times$ 10$^{-5}$ & \multirow{2}{*}{0.6} & \multirow{2}{*}{30.2/($44-4$)} & 2.14 $\times$ 10$^{-4}$\\
      $Y(4230)$$^{II}$ & 2.70 $\pm$ 0.39 & 3.53 $\pm$ 0.05 & 4.68 $\pm$ 0.01 & (10.10 $\pm$ 0.12) $\times$ 10$^{-1}$ & &  & $\ldots$\\
      \hline    
      $Y(4360)$$^I$ & 2.65 $\pm$ 0.41 & 3.52 $\pm$ 0.05 & 5.16 $\pm$ 0.85 & (1.48 $\pm$ 2.70) $\times$ 10$^{-4}$ & \multirow{2}{*}{0.6} & \multirow{2}{*}{30.1/($44-4$)} & 4.95 $\times$ 10$^{-4}$\\
      $Y(4360)$$^{II}$ & 2.67 $\pm$ 0.41 & 3.52 $\pm$ 0.05 & 4.67 $\pm$ 0.01 & 2.02 $\pm$ 0.02 & & & $\ldots$\\
      \hline    
      $\psi$(4415)$^I$ & 2.76 $\pm$ 0.41 & 3.53 $\pm$ 0.05 & 5.95 $\pm$ 0.66 & (2.60 $\pm$ 3.80) $\times$ 10$^{-4}$ & \multirow{2}{*}{0.9} & \multirow{2}{*}{29.4/($44-4$)} & 7.47 $\times$ 10$^{-4}$\\    
      $\psi$(4415)$^{II}$ & 2.77 $\pm$ 0.42 & 3.53 $\pm$ 0.05 & 4.68 $\pm$ 0.01 & 1.72 $\pm$ 0.02 & & & $\ldots$ \\  
      \hline
      $Y(4500)$$^I$ & 3.08 $\pm$ 0.48 & 3.57 $\pm$ 0.06 & 0.85 $\pm$ 0.37 & (1.33 $\pm$ 1.50) $\times$ 10$^{-3}$ & \multirow{2}{*}{1.4} & \multirow{2}{*}{27.6/($44-4$)} & 3.25 $\times$ 10$^{-3}$\\    
      $Y(4500)$$^{II}$ & 3.10 $\pm$ 0.49 & 3.58 $\pm$ 0.06 & 4.69 $\pm$ 0.01 & 1.67 $\pm$ 0.04 & & & $\ldots$\\   
      \hline
      $Y(4710)$$^I$ & 3.42 $\pm$ 0.61 & 3.62 $\pm$ 0.06 & 2.75 $\pm$ 0.27 & (3.37 $\pm$ 2.36) $\times$ 10$^{-3}$ & \multirow{2}{*}{2.4} & \multirow{2}{*}{22.9/($44-4$)} & 6.40 $\times$ 10$^{-3}$\\    
      $Y(4710)$$^{II}$ & 3.46 $\pm$ 0.62 & 3.62 $\pm$ 0.06 & 4.60 $\pm$ 0.02 & 2.03 $\pm$ 0.04 & & & $\ldots$\\ 
      \hline\hline
    \end{tabular}    
    }   
    \label{tab:XYZ_result}
  \end{table*}

  \begin{figure*}[htbp]
    \centering
    \includegraphics[width=0.5\textwidth]{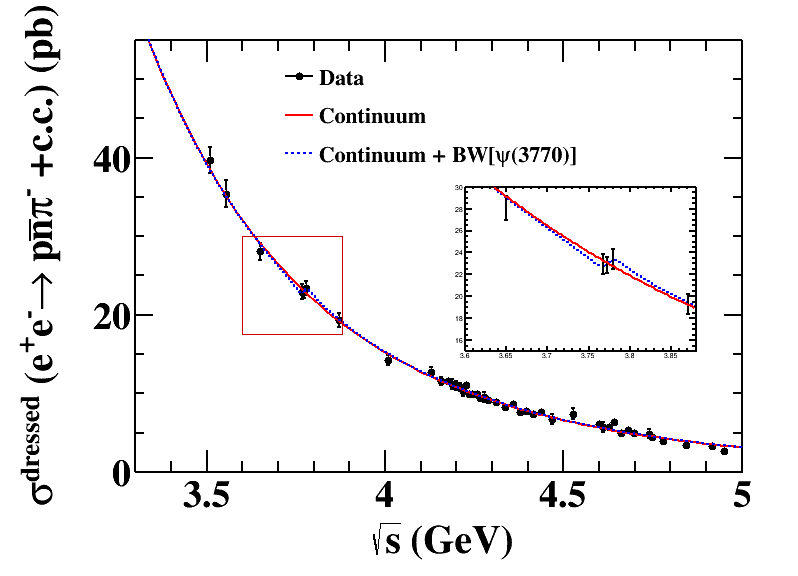}%
    \includegraphics[width=0.5\textwidth]{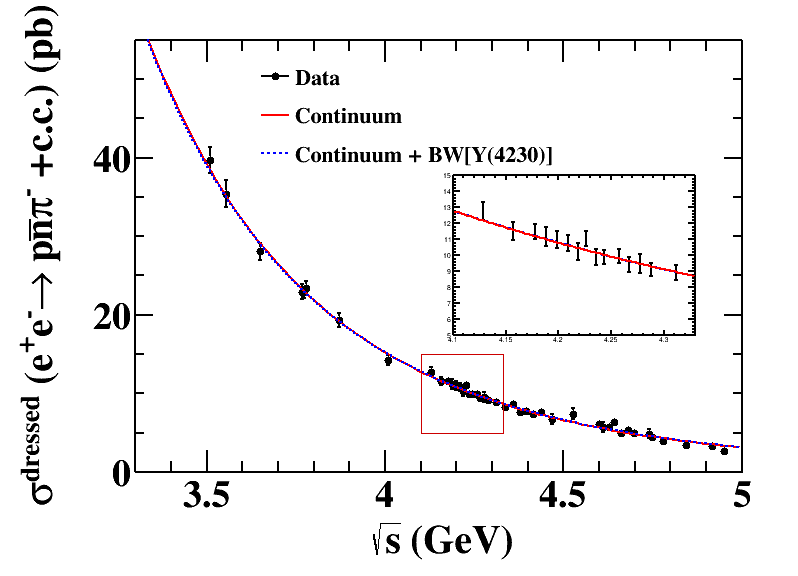}\\%
    \includegraphics[width=0.5\textwidth]{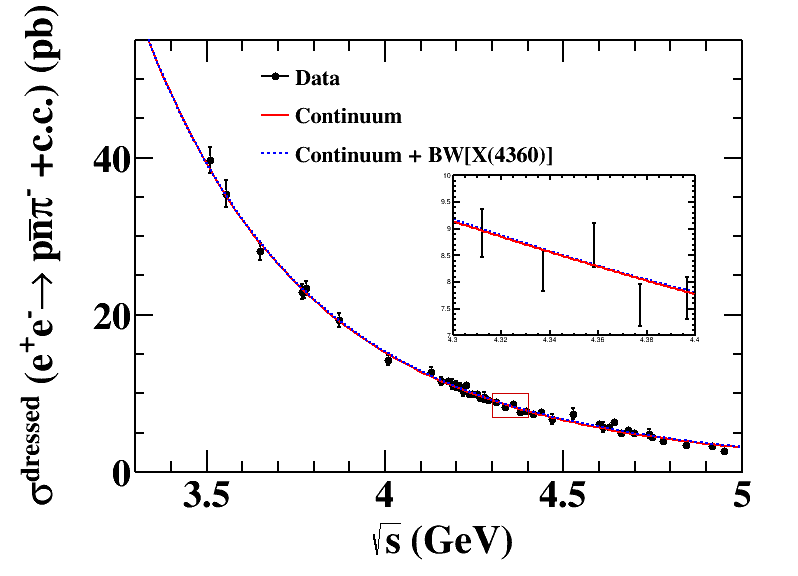}%
    \includegraphics[width=0.5\textwidth]{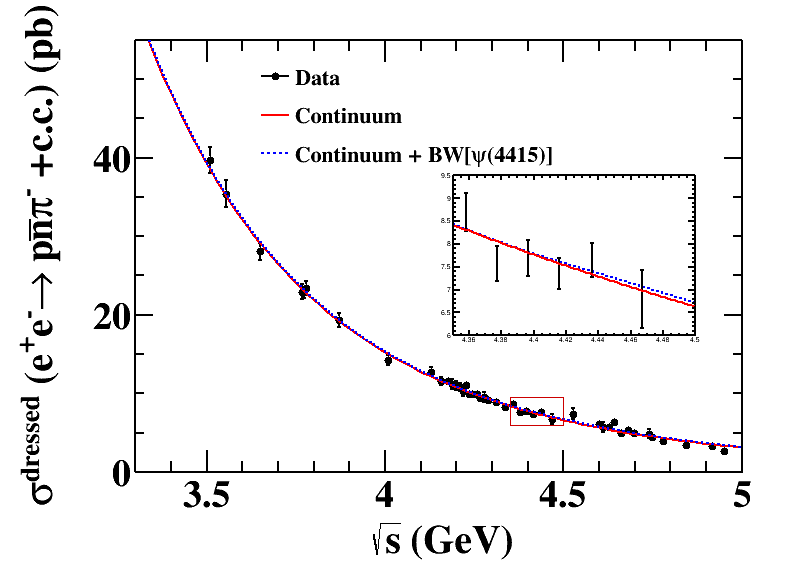}\\%
    \includegraphics[width=0.5\textwidth]{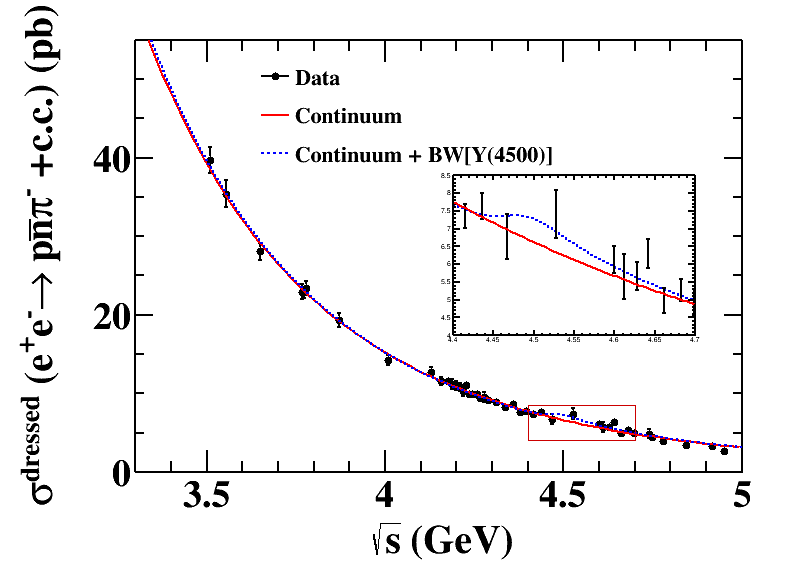}%
    \includegraphics[width=0.5\textwidth]{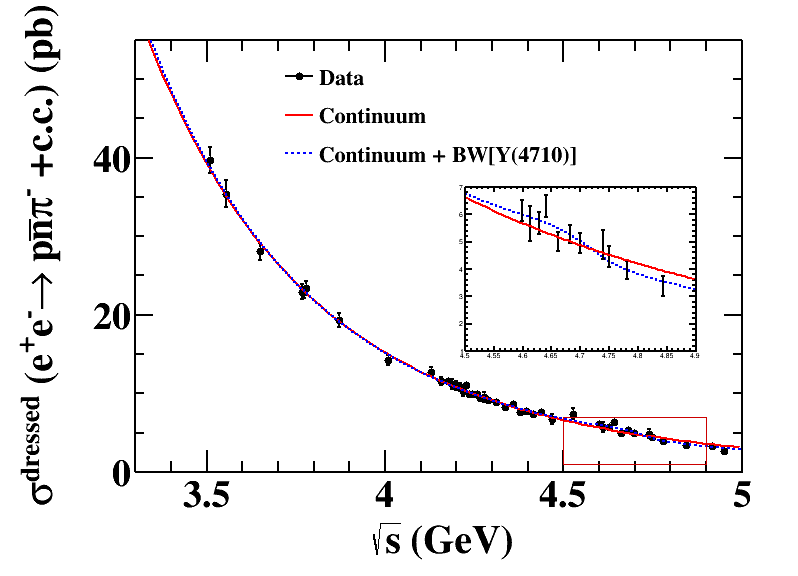}\\%
    \caption{Fits to $\sigma^{\mathrm{dressed}}$($e^{+}e^{-}\to p \bar{n} \pi^{-} + \mathrm{c.c.}$) with resonance ($\psi(3770)$, $Y(4230)$, $X(4360)$, $\psi(4415)$, $Y(4500)$, or $Y(4710)$) and continuum (dashed line), or only continuum term (solid line). Dots with error bars are the measured dressed cross sections. The uncertainties are statistical and systematic. The insets show an enlarged view of the respective resonance mass regions to demonstrate the effect of including the resonance state in the fit.}  
    \label{fit_Born2}
  \end{figure*}

  \section{\label{sec:fit} FIT TO THE dressed CROSS SECTIONS}

  Since the processes $e^{+}e^{-}\to p\bar{n}\pi^{-}$ and $e^{+}e^{-}\to \bar{p}n\pi^{+}$ 
  are charge-conjugate channels, their dressed cross sections are expected to be 
  identical. Experimentally, the ratio of the measured cross sections is found to be 
  $1.02 \pm 0.01$, consistent with unity within uncertainties, as shown in 
  Fig.~\ref{fig:com_cs_pnpi}. The pull distribution shown in the bottom panel indicates 
  good agreement between data and fit. Therefore, in this work the cross sections 
  measured from the two conjugate channels are combined and used to fit the line shape 
  in order to investigate possible contributions from resonant states. The total dressed 
  cross section is obtained by summing the two charge-conjugate channels as
  \begin{equation}  
    \sigma^{\mathrm{dressed}}_{\mathrm{tot}} = \sigma^{\mathrm{dressed}}_{p\bar{n}\pi^-} + \sigma^{\mathrm{dressed}}_{\bar{p}n\pi^+}, 
  \end{equation}
  where $\sigma^{\mathrm{dressed}}_{p\bar{n}\pi^-}$ and $\sigma^{\mathrm{dressed}}_{\bar{p}n\pi^+}$ represent the dressed cross sections measured for the processes 
  $e^{+}e^{-}\to p\bar{n}\pi^{-}$ and $e^{+}e^{-}\to \bar{p}n\pi^{+}$, respectively. 
  To calculate the corresponding total uncertainty of the sum, the independent 
  statistical and systematic uncertainties of the individual channels are added in 
  quadrature. The fully correlated systematic uncertainties between the two channels, 
  such as those originating from the luminosity and the MC model, are combined linearly 
  before being added in quadrature with the independent components. 
  Figure~\ref{fit_Born2} shows the combined dressed cross sections of the two 
  charge-conjugate channels measured in the considered energy range. 

  Potential resonances in the line shape of the cross sections for the process 
  $e^+e^- \to p \bar{n} \pi^{-}$ are studied by fitting the dressed cross sections 
  $\sigma^{\text{dressed}}$ = $\frac{\sigma^{B}}{{|1 - \Pi|}^2}$. Assuming that the 
  $e^+e^- \to p \bar{n} \pi^{-}$ process is produced through both resonance decay and 
  continuum processes, the dressed cross sections are fitted using the coherent sum of 
  a power-law (PL) function and a Breit-Wigner (BW) function, defined as
  \begin{equation}\label{XYZ}
    \sigma(\sqrt{s}) = \left| \sqrt{\sigma_{\text{con}}} + \mathrm{BW}(\sqrt{s})\, e^{i\phi} \right|^2 ,
  \end{equation}
  where $\sigma_{\text{con}}$ denotes the continuum contribution, parameterized as a 
  function of $s$ by $\sigma_{\text{con}} = \frac{C}{s^{\lambda}}$, with $s$ being the 
  square of the c.m.\ energy, and the parameters $C$ and $\lambda$ being free parameters 
  determined from the fit. The term $\mathrm{BW}(\sqrt{s})$ represents the BW amplitude 
  in the narrow-width approximation, $\mathrm{BW}(\sqrt{s}) = \frac{\sqrt{12\pi\,\Gamma_{e^+e^-}\,\Gamma\,\mathcal{B}(\psi \to p\bar{n}\pi^-)}}{s - M^2 + iM\Gamma}$, 
  which describes the resonant contribution characterized by the resonance mass $M$, 
  total width $\Gamma$, electronic partial width $\Gamma_{e^+e^-}$, and branching 
  fraction $\mathcal{B}(\psi \to p\bar{n}\pi^-)$. The parameter $\phi$ describes the 
  phase between the resonant and continuum production amplitudes. The mass $M$ and 
  width $\Gamma$ of the resonance $\psi$ are fixed to the PDG values~\cite{ParticleDataGroup:2024cfk}. The values of $C$, $\lambda$, $\Gamma_{e^+ e^-}\times\mathcal{B}$, 
  and the interference phase $\phi$ are free in the fit.

  To investigate the possible contribution of the $\psi(3770)$ resonance to the process 
  $e^{+}e^{-}\to p \bar{n} \pi^{-}$, the data are fitted using Eq.~\eqref{XYZ}; the 
  results are shown in Fig.~\ref{fit_Born2}. No significant resonant structure is 
  observed. The fit parameters are presented in Table~\ref{tab:XYZ_result}. The 
  statistical significance is calculated based on the change in the $\chi^2$ value and 
  the number of free parameters in the fits with and without the $\psi(3770)$ resonance. 
  The best fit yields two distinct solutions with identical 
  $\chi^2/\text{NDF} = 31.5/40$, corresponding to equivalent fit quality. A detailed 
  explanation and mathematical review of the multiple-solution problem is given in 
  Ref.~\cite{zhu:2011ha}. The solutions for $\Gamma_{e^+ e^-} \times \mathcal{B}$ are 
  $(2.31 \pm 5.06) \times 10^{-4}$~eV (solution I) and 
  $(9.46 \pm 0.12) \times 10^{-1}$~eV (solution II), respectively. The results for the 
  phases between the resonant and continuum production amplitudes are 
  $0.05 \pm 0.88$~rad (solution I) and $4.72 \pm 0.02$~rad (solution II).  

  Due to the large uncertainty of $\Gamma_{e^+ e^-} \times \mathcal{B}$, the upper limit 
  at the 90\% confidence level (C.L.) is given (assuming that the uncertainty follows a 
  Gaussian distribution). The upper limit on $\Gamma_{e^+ e^-} \times \mathcal{B}$ at 
  the 90\% C.L., $({\Gamma_{e^+ e^-} \times \mathcal{B}})^{\text{up}}$, is determined 
  by
  \[
    \frac{
      \displaystyle \int_0^{(\Gamma_{e^+ e^-} \times \mathcal{B})^{\text{up}}}
      \mathrm{G}\!\left(\Gamma_{e^+ e^-} \times \mathcal{B}, \, \delta_{\Gamma_{e^+ e^-} \times \mathcal{B}}\right) \, dx
      }{
        \displaystyle \int_0^{\infty}
        \mathrm{G}\!\left(\Gamma_{e^+ e^-} \times \mathcal{B}, \, \delta_{\Gamma_{e^+ e^-} \times \mathcal{B}}\right) \, dx
        }
        = 0.9,
      \]
      where \( \mathrm{G}(\Gamma_{e^+ e^-} \times \mathcal{B}, \delta_{\Gamma_{e^+ e^-} \times \mathcal{B}}) \) is a Gaussian function with mean 
      value \( \Gamma_{e^+ e^-} \times \mathcal{B} = 2.31 \times 10^{-4} \,~\text{eV} \) 
      and standard deviation \( \delta_{\Gamma_{e^+ e^-} \times \mathcal{B}} = 5.06 \times 10^{-4}\,~\text{eV} \). The obtained upper limit is 8.79 $\times$ 10$^{-4}$~eV.

      \begin{figure*}[!htbp]  
        \centering  
        \includegraphics[width=1.0\linewidth]{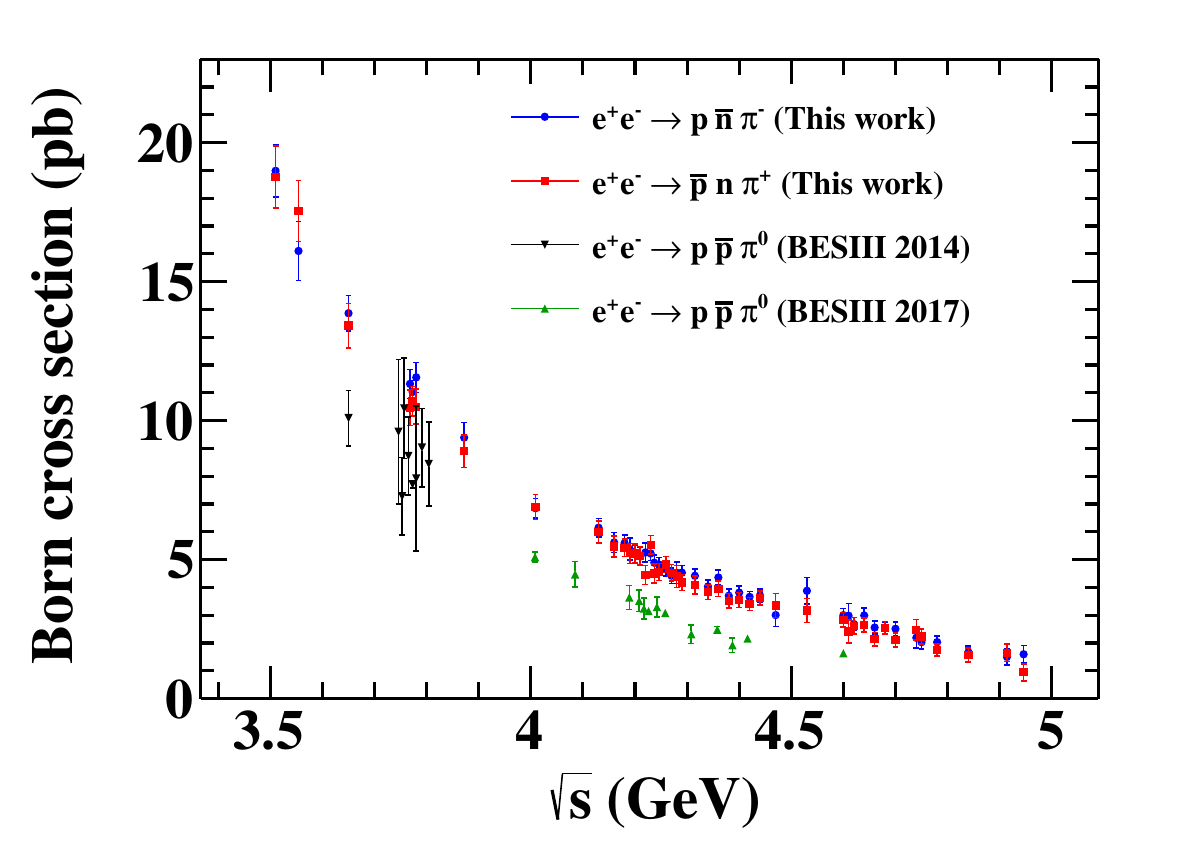}  
        \caption{Measurement of the Born cross sections for $e^{+}e^{-}\to p\bar{n}\pi^{-}$ and comparison with $e^{+}e^{-}\to p\bar{p}\pi^{0}$~\cite{BESIII:2014lpf,BESIII:2017qwj}.}  
        \label{fig:placeholder}
      \end{figure*}

      In the energy range $\sqrt{s} = 3.510$--$4.946$~$\GeV$, the measured cross 
      sections are well described by the coherent interference between the resonant and 
      continuum amplitudes. The resonance may correspond to the well-established 
      $\psi(3770)$, or to one of the higher-mass charmonium-like states, such as 
      $Y(4230)$, $Y(4360)$, $\psi(4415)$, $Y(4500)$~\cite{BESIII:2022joj}, 
      or $Y(4710)$~\cite{BESIII:2022kcv}. Table~\ref{tab:XYZ_result} summarizes 
      the fit results. None of the assumed resonances is observed with a significance 
      exceeding $3\sigma$ in the process $e^+e^- \to p \bar{n} \pi^{-}$. Consequently, 
      the upper limits on the products of the branching fraction and the electronic 
      partial width are determined at the 90\% C.L. The fit results for the states 
      $Y(4230)$, $Y(4360)$, $\psi(4415)$, $Y(4500)$~\cite{BESIII:2022joj}, and 
      $Y(4710)$~\cite{BESIII:2022kcv} are shown in Fig.~\ref{fit_Born2}. Zoomed-in 
      insets are provided in the figure around the mass region of each assumed 
      resonance to highlight the detailed fit variations when the resonance is included.

      \section{\label{sec:sum} Summary}
      The Born cross sections for the process $e^+e^- \to p \bar{n} \pi^-$ are measured 
      at 44 c.m.\ energies between 3.510 and 4.946~$\GeV$. The measured cross sections 
      decrease with increasing $\sqrt{s}$, and the ratio of the cross sections for 
      the $p\bar{n}\pi^-$ and $\bar{p}n\pi^+$ channels is determined to be $1.02\pm0.01$.
      No significant resonant structures associated with $\psi(3770)$, $Y(4230)$, 
      $Y(4360)$, $\psi(4415)$, $\psi(4500)$, or $\psi(4710)$ are observed, and the upper 
      limits at the 90\% C.L.\ on the products of the branching fractions and the 
      electronic partial widths are obtained. The measured cross sections for 
      $e^+e^- \to p \bar{n} \pi^-$ consistently exceed those for 
      $e^+e^- \to p \bar{p} \pi^0$~\cite{BESIII:2014lpf, BESIII:2017qwj}. 
      Although isospin symmetry predicts an exact 2:1 ratio for the decay of a purely 
      isoscalar resonance, our lineshape analysis reveals no significant resonance 
      contributions, indicating that continuum processes dominate the production. 
      The observed enhancement in the charged pion channel can be attributed to two 
      primary continuum mechanisms. First, intermediate two-body decays involving either 
      baryon ($N^*$) or meson ($\rho^*$) resonances naturally favor charged pions over 
      neutral ones by a factor of two, as dictated by isospin Clebsch-Gordan 
      coefficients. Second, in direct three-body production, the charged pion may 
      couple directly to the virtual photon, thereby yielding an enhanced production 
      ratio~\cite{Gakh:2022fad}. To quantitatively determine the fractional 
      contributions of these intermediate processes and fully clarify the production 
      dynamics, a dedicated partial wave analysis with higher statistics is required. 
      Such future investigations will yield valuable insights into the role of 
      intermediate resonances and isospin-related effects in baryon-antibaryon production.

      \input{acknowledgement_2026-04-21}

      \bibliography{refs}
\end{document}

%% file: authorlist_2026-04-21.tex
M.~Ablikim$^{1}$\BESIIIorcid{0000-0002-3935-619X},
M.~N.~Achasov$^{4,c}$\BESIIIorcid{0000-0002-9400-8622},
P.~Adlarson$^{84}$\BESIIIorcid{0000-0001-6280-3851},
X.~C.~Ai$^{90}$\BESIIIorcid{0000-0003-3856-2415},
C.~S.~Akondi$^{32A,32B}$\BESIIIorcid{0000-0001-6303-5217},
R.~Aliberti$^{40}$\BESIIIorcid{0000-0003-3500-4012},
A.~Amoroso$^{83A,83C}$\BESIIIorcid{0000-0002-3095-8610},
Q.~An$^{79,66,\dagger}$,
Y.~H.~An$^{90}$\BESIIIorcid{0009-0008-3419-0849},
M.~S.~Anderson$^{40}$\BESIIIorcid{0009-0008-1550-2632}
% ...
\\
\vspace{0.2cm}
(BESIII Collaboration)\\
\vspace{0.2cm} {\it
$^{1}$ Institute of High Energy Physics, Beijing 100049, People's Republic of China\\
% ...
$^{4}$ Budker Institute of Nuclear Physics SB RAS (BINP), Novosibirsk 630090, Russia\\
% ...
$^{32}$ INFN Sezione di Ferrara, (A)INFN Sezione di Ferrara, I-44122, Ferrara, Italy; (B)University of Ferrara, I-44122, Ferrara, Italy\\
% ...
$^{40}$ Johannes Gutenberg University of Mainz, Johann-Joachim-Becher-Weg 45, D-55099 Mainz, Germany\\
% ...
$^{66}$ State Key Laboratory of Particle Detection and Electronics, Beijing 100049, Hefei 230026, People's Republic of China\\
% ...
$^{79}$ University of Science and Technology of China, Hefei 230026, People's Republic of China\\
% ...
$^{83}$ University of Turin and INFN, (A)University of Turin, I-10125, Turin, Italy; (B)University of Eastern Piedmont, I-15121, Alessandria, Italy; (C)INFN, I-10125, Turin, Italy\\
$^{84}$ Uppsala University, Box 516, SE-75120 Uppsala, Sweden\\
% ...
$^{90}$ Zhengzhou University, Zhengzhou 450001, People's Republic of China\\

\vspace{0.2cm}
$^{\dagger}$ Deceased\\
% ...
$^{c}$ Also at the Novosibirsk State University, Novosibirsk, 630090, Russia\\
% ...
}

%% file: acknowledgement_2026-04-21.tex
\textbf{Acknowledgement}

The BESIII Collaboration thanks the staff of BEPCII (https://cstr.cn/31109.02.BEPC) and the IHEP computing center for their strong support. This work is supported in part by National Key R\&D Program of China under Contracts Nos. 2023YFA1606704,
2023YFA1606000, 2025YFA1613900; National Natural Science Foundation of China (NSFC) under Contracts Nos. 12275320, 12475077, 11635010, 11935015, 11935016, 11935018, 12025502, 12035009, 12035013, 12061131003, 12192260, 12192261, 12192262, 12192263, 12192264, 12192265, 12221005, 12225509, 12235017, 12342502, 12361141819, 12535005; the Chinese Academy of Sciences (CAS) Large-Scale Scientific Facility Program; the Strategic Priority Research Program of Chinese Academy of Sciences under Contract No. XDA0480600; CAS under Contract No. YSBR-101; 100 Talents Program of CAS; The Institute of Nuclear and Particle Physics (INPAC) and Shanghai Key Laboratory for Particle Physics and Cosmology; Agencia Nacional de Investigación y Desarrollo de Chile (ANID), Chile under Contract No. ANID CCTVal CIA250027; ERC under Contract No. 758462; German Research Foundation DFG under Contract No. FOR5327; Istituto Nazionale di Fisica Nucleare, Italy; Knut and Alice Wallenberg Foundation under Contracts Nos. 2021.0174, 2021.0299, 2023.0315; Ministry of Development of Turkey under Contract No. DPT2006K-120470; National Research Foundation of Korea under Contract No. RS-2026-25486791; National Science and Technology fund of Mongolia; Polish National Science Centre under Contract No. 2024/53/B/ST2/00975; STFC (United Kingdom); Swedish Research Council under Contract No. 2019.04595; U. S. Department of Energy under Contract No. DE-FG02-05ER41374.